\documentclass[twocolumn,twocolappendix]{aastex631}
\usepackage[utf8]{inputenc}
\usepackage{newtxtext,newtxmath}
\usepackage[T1]{fontenc}
\usepackage{ae,aecompl}
\usepackage{comment}

\usepackage{graphicx,tabularx}	
\usepackage{amsmath}	
\usepackage{amssymb}	
\usepackage{booktabs}
\usepackage[inline,shortlabels]{enumitem}
\setlist{itemjoin* = { and\enspace}}
\usepackage{color,soul,xcolor}
\usepackage{multirow}

\usepackage[figuresright]{rotating}

\usepackage{threeparttable}

\newcommand{\mz}[1]{\textbf{\textcolor{blue}{[M:#1]}}}
\newcommand{\al}[1]{\textbf{\textcolor{magenta}{[A:#1]}}}

\received{XXX}
\revised{XXX}
\accepted{XXX}

\submitjournal{ApJ}

\shorttitle{${\rm kSZ^2}$-${\rm 21cm^2}$ Cross-Correlations during Reionization}
\shortauthors{Zhou, et al.}

\graphicspath{{./}{figures/}}
\usepackage{tablefootnote}
\begin{document}

\title{Prospects for ${\rm kSZ^2}$-${\rm 21cm^2}$ Cross-Correlations during Reionization}

\correspondingauthor{Meng Zhou}
\email{zhoumeng@bao.ac.cn}

\author[0000-0002-2744-0618]{Meng Zhou}
\affiliation{National Astronomical Observatories, Chinese Academy of Sciences, 20A Datun Road, Bejing 100101, P. R. China}

\author[0000-0002-4693-0102]{Paul La Plante}
\affiliation{Department of Computer Science, University of Nevada, Las Vegas, NV 89154, USA}
\affiliation{Nevada Center for Astrophysics, University of Nevada, Las Vegas, NV 89154, USA}

\author[0000-0002-3950-9598]{Adam Lidz}
\affiliation{Center for Particle Cosmology, Department of Physics and Astronomy, University of Pennsylvania, Philadelphia, PA 19104, USA}

\author[0000-0002-1301-3893]{Yi Mao} 
\affiliation{Department of Astronomy, Tsinghua University, Beijing 100084, P. R. China}

\author[0000-0001-8108-0986]{Yin-Zhe Ma}
\affiliation{Department of Physics, Stellenbosch University, Matieland 7602, South Africa}
\affiliation{National Institute for Theoretical and Computational Science (NITheCS), Stellenbosch, Matieland 7602, South Africa}



\begin{abstract}

The 21\,cm line and the patchy kinetic Sunyaev-Zel'dovich (kSZ) effect are promising and complementary probes of the Epoch of Reionization (EoR). A challenge for cross-correlating these two signals is that foreground avoidance or removal algorithms applied to the 21\,cm data inevitably sacrifice Fourier modes with long wavelengths along the line-of-sight (i.e., low-$k_\parallel$ modes), yet \textit{only} these same modes contribute to the kSZ signal. Here we show that a suitable kSZ$^2$ $\times$ 21\,cm$^2$ cross-correlation statistic nevertheless remains non-vanishing, even after filtering out the corrupted low-$k_\parallel$ Fourier modes from the 21\,cm data. We simulate the kSZ$^2$ $\times$ 21\,cm$^2$ cross-correlation signal across reionization-era redshifts and find distinctive redshift evolution. This signal peaks early in the reionization history, when the volume-averaged fraction is around $0.1 \lesssim x_\mathrm{HII} \lesssim 0.2$, after which it changes sign and reaches a minimum near reionization's midpoint ($x_\mathrm{HII} \sim 0.5$), while the signal gradually vanishes as reionization completes. These trends appear generic across three simulated models which differ in their reionization histories. We forecast the detectability of the kSZ$^2$ $\times$ 21\,cm$^2$ cross-power spectrum for the HERA and SKA1-Low 21\,cm experiments in combination with current and next-generation CMB surveys including the Simons Observatory, CMB-S4, and CMB-HD. We find that a high-significance detection ($\mathrm{S/N} \gtrsim 5\sigma$) is possible with SKA1-Low and CMB-S4.

\end{abstract}



\section{Introduction} \label{sec:intro}
The Epoch of Reionization is among the least explored periods in the history of our universe. 
During the EoR, early generations of stars and galaxies formed at $5 \lesssim z \lesssim 20$, emitted ultraviolet (UV) photons and ionized surrounding neutral hydrogen, producing a patchy network of ionized bubbles.  
These bubbles grew, merged, and finally filled the entire universe. Since cosmic reionization should be driven by the photons produced by early galaxies, measurements of the spatial fluctuations during the EoR can be considered indirect high-redshift galaxy surveys. Therefore, research into the EoR will help us understand the properties of early galaxies and their formation mechanisms. 

The 21\,cm line from the hyperfine transition of neutral hydrogen (H~{\small I}) is one of the most promising probes of the EoR because it traces spatial fluctuations in the density of H~{\small I}. Observationally, radio interferometric measurements are the most promising approach for detecting the 21\,cm signal from the EoR. Over the last couple of decades, there have been several attempts to measure the 21\,cm power spectrum (e.g. LOFAR, \citealt{2013A&A...556A...2V}; MWA, \citealt{2013PASA...30....7T}; and PAPER, \citealt{2010AJ....139.1468P}, among others). More sensitive experiments are also ongoing (e,g. HERA, \citealt{2017PASP..129d5001D} and the SKA, \citealt{2009IEEEP..97.1482D}). These experiments must overcome many technical and observational hurdles before they will be able to detect the 21\,cm power spectrum from the EoR at high statistical significance. The main challenges owe to foreground contamination from synchrotron radiation and free-fee emission from our Milky Way and extragalactic sources, which can exceed the cosmological signal by 4-5 orders of magnitude. Other systematic challenges, which couple to bright foregrounds, include calibration errors, the chromatic response of an interferometer, and radio frequency interference.

Another potentially detectable signal from the EoR is the kinetic Sunyaev-Zel'dovich (kSZ) effect due to the inverse Compton scattering of CMB photons by free electrons. The free electrons during reionization are expected to show a patchy distribution, tracing the ionized bubbles, and so the reionization-era contribution to the kSZ signal is referred to as the ``patchy kSZ effect''. 
The total kSZ signal includes both patchy reionization contributions and those from ionized gas in the post-reionization universe. This signal has been detected in current surveys at roughly 3$\sigma$ statistical significance via an angular power spectrum analysis \citep{2014JCAP...04..014D,2021ApJ...908..199R}. An upcoming suite of wide-field, high-sensitivity, and fine-angular-resolution CMB surveys, including the Simons Observatory (SO, \citealt{2019JCAP...02..056A}), CMB-S4 \citep{2016arXiv161002743A}, and CMB-HD \citep{2019BAAS...51g...6S}, also seek to measure the kSZ signal to high significance. 

However, it is difficult to extract the patchy kSZ signal from CMB surveys and to exploit it as a constraint on reionization. At the relevant angular scales and observational frequencies, the kSZ-induced anisotropies need to be carefully separated from other contributions, including the primary CMB anisotropies, fluctuations in the cosmic infrared background (CIB) and those sourced by the thermal Sunyaev-Zel'dovich (tSZ) effect. Furthermore, one must also separate the EoR component to the kSZ effect from the post-reionization contribution, the so-called ``late-time kSZ'' signal. It is therefore challenging to separate the reionization-era kSZ signal from other contributions to the CMB anisotropies. 

One naive idea for bypassing these difficulties is to perform a two-point cross-correlation between the kSZ signal and the 21\,cm line, or any other reionization-era large-scale structure (LSS) tracer. Unfortunately, a new challenge arises in such cross-correlations with the kSZ signal: a cancellation occurs because ionized regions may be either moving towards or away from the observer \citep{2004ApJ...606...46D}. One possible solution to this issue is to filter the CMB map to suppress contamination from other components, and then square the filtered map \citep{2004ApJ...606...46D,2016PhRvD..94l3526F,2016PhRvL.117e1301H,2022ApJ...928..162L}. The cross-correlation signal can then be computed between this filtered-and-squared map and the preferred LSS tracer. Although several previous studies have looked at the prospects for measuring the cross-power spectrum between the 21\,cm and kSZ fields during reionization \citep{2006ApJ...647..840A,2018MNRAS.476.4025M,2020ApJ...899...40L}, foreground contamination in the 21\,cm surveys at low-$k_\parallel$ will swamp any joint signal with the kSZ.

In principle, one could just choose an alternate tracer to combine with the kSZ signal without the foreground contamination/low-$k_\parallel$ mode issue, e.g. galaxies \citep{2022ApJ...928..162L}. Considering that the 21\,cm line is one of the most promising tracers of the EoR, it is nevertheless worth developing a new method to suppress foreground contamination in cross-correlation measurements between the kSZ signal and the 21\,cm line. 

Here we study, for the first time, the kSZ$^2$--21\,cm$^2$ cross-correlation during the EoR. In addition to filtering and squaring the kSZ signal, we will apply a related procedure to the 21\,cm field as well. 
For the 21\,cm field, we aim to filter out the modes contaminated by foreground emission in Fourier space and then square the resulting field in configuration space. The filtered and then squared field is a convolution in Fourier space: as such,  at low $k_\parallel$, it still includes contributions from pairs of 21 cm modes with large line-of-sight wavenumbers, provided the line-of-sight components are equal and opposite (see Appendix~\ref{appendix_a} for the explicit details here). Hence, the filtered and squared 21\,cm field may be cross-correlated with the squared kSZ signal, even after filtering the foreground-corrupted low $k_\parallel$ modes from the 21\,cm data cube. In this case, the resulting 21\,cm$^2$ signal has some underlying information in common with the reionization-era kSZ$^2$ signal, and this can potentially be extracted with a kSZ$^2$--21\,cm$^2$ cross-power spectrum analysis.    
This investigation extends work in the current literature, which explored related higher-order statistics including 21\,cm$^2$--NIRB cross-correlations \citep{2024arXiv241021410S},  21\,cm--21\,cm--CMB lensing estimators \citep{2023arXiv231105904M}, and kSZ 4-point function statistics \citep{2017PhRvL.119b1301S,2018PhRvD..98l3519F}. Here, we employ reionization simulations to model the kSZ$^2$--21\,cm$^2$ cross-power spectrum signal, and forecast its detectability in various current and next-generation experiments.  

The spatial average of the filtered and squared kSZ signal has a relatively simple interpretation: it measures the kSZ angular power spectrum integrated over the multipoles which are extracted by the filter. The spatially-averaged filtered and squared 21 cm field has a similar interpretation, except that it measures the 21 cm fluctuation power in particular redshift chunks, which can be varied in the analysis. The cross-power spectrum between the two fields, kSZ$^2$--21\,cm$^2$, then essentially measures the covariance between the 21 cm power spectrum and the kSZ power spectrum in different multipole bands and redshift chunks. More precisely, this cross-power spectrum may be expressed as an integral over an electron momentum density--electron momentum density-- 21\,cm--21\,cm trispectrum (see Appendix~\ref{appendix_b}). 

This paper is organized as follows. In Section~\ref{sec: method}, we describe the semi-numeric simulation methods used to model reionization, the kSZ, and 21\,cm signals. In Section~\ref{sec: results}, we present the simulated kSZ$^2$--21\,cm$^2$ cross-power spectrum and discuss its dependence on the ionization history and the filters employed. In Section~\ref{sec: detectability}, we estimate the feasibility of detecting this signal from upcoming kSZ and 21\,cm observations. Finally, in Section~\ref{sec: conclusion} we draw our conclusions.

\section{Methods} \label{sec: method}
In this work, we use the mock 21\,cm and kSZ data generated in \citet{2022ApJ...928..162L} and \citet{2023ApJ...944...59L}. In this section, we first review how these mock data are generated. Then we describe the pipeline we use for calculating the angular cross-power spectrum.
\subsection{Reionization Simulation}
\label{sec: simulation}

We generate mock reionization data using {\tt zreion} \citep{2013ApJ...776...81B}, a semi-numeric reionization method which can be applied to the kSZ and 21 cm signals, as well as to other tracers of cosmic reionization. This technique determines whether each simulated grid cell is ionized or neutral at given redshift values, with the ionization state approximated as completely ionized or completely neutral in each cell. Its basic assumption is that the ``reionization overdensity field'' $\delta_z$ is a biased tracer of the matter overdensity field $\delta_m$ on large scales:
\begin{equation}
   \tilde{\delta}_z(\vec{\mathbf{k}}) = b_{zm}\tilde{\delta}_m(\vec{\mathbf{k}}).
\end{equation}
Here, $\tilde{\delta}_z(\vec{\mathbf{k}})$ and $\tilde{\delta}_m(\vec{\mathbf{k}})$ are the Fourier transforms of $\delta_z(\vec{\mathbf{r}})$ and $\delta_m(\vec{\mathbf{r}})$, respectively. The reionization overdensity field $\delta_z(\vec{\mathbf{r}})$ is defined as:
\begin{equation}
    \delta_z(\vec{\mathbf{r}}) = \frac{[1+z_{\rm RE}(\vec{\mathbf{r}})]-[1+\bar{z}]}{1+\bar{z}}\,,
\end{equation}
where $z_{\rm RE}(\vec{\mathbf{r}})$ is the redshift at which the grid cell at position $\vec{\mathbf{r}}$ is first ionized, and $\bar{z}$ is the midpoint of reionization. 

\begin{figure}[!h]
    \centering
    \includegraphics[width=0.9\columnwidth]{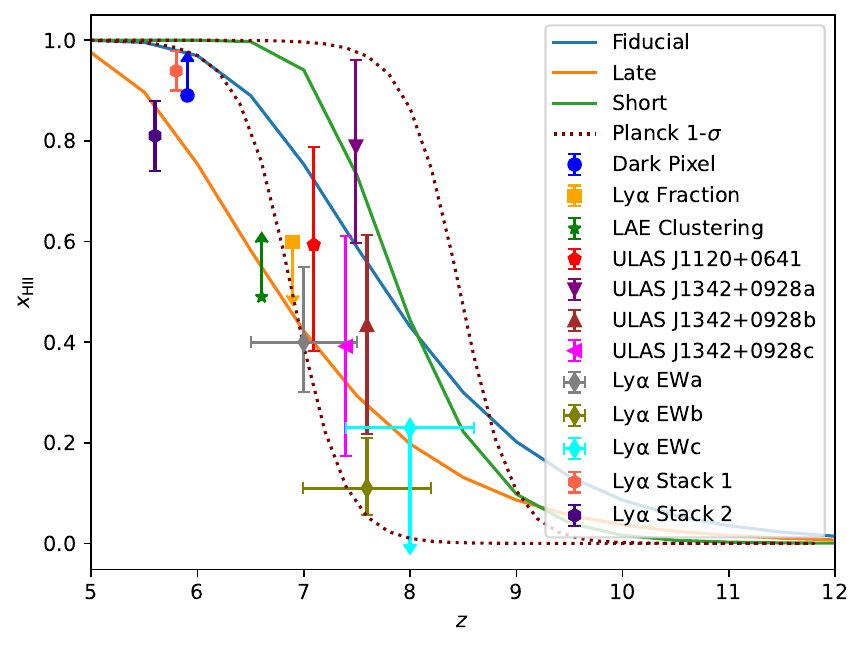}
    \caption{The global ionization fraction, $x_{\rm HII}$, as a function of redshift in our fiducial, late, and short reionization scenarios. We also show observational constraints from dark pixel fraction measurements, Ly$\alpha$ fraction evolution, the clustering of Ly$\alpha$ emitters, QSO damping wings, Ly$\alpha$ equivalent width distribution, Ly$\alpha$ forest stacking around highly absorbed regions, and the inferred TANH model from Planck. All three reionization scenarios are consistent with these observational constraints.}
    \label{fig:hist}
\end{figure}

In principle, one can fit the bias function to any arbitrary form based on the chosen benchmark simulation. Here the bias is parameterized in the following way:
\begin{equation}
\label{eqn:bias}
    b_{zm}(k) = \frac{b_0}{\left(1+\frac{k}{k_0}\right)^\alpha}\,,
\end{equation}
where $b_0 = 1/\delta_c = 0.593$, and $\delta_c$ is the critical overdensity according to linear theory. Therefore, the global history of reionization is completely determined by the parameters $\{\bar{z},\ \alpha,\ k_0\}$. We list the values of the parameters in each simulated model considered in this paper in Table~\ref{tab:simulation_para} and plot the corresponding reionization history in Figure~\ref{fig:hist}. We also include recent observational constraints on the reionization history from: using the fraction of dark pixels in the Ly-a forest\citep{2015MNRAS.447..499M}, Ly$\alpha$ fraction evolution\citep{2015MNRAS.446..566M}, 
clustering of Ly$\alpha$ emitters\citep{2010ApJ...723..869O,2015MNRAS.453.1843S}, QSO damping wings \citep{2017MNRAS.466.4239G,2019MNRAS.484.5094G,2018Natur.553..473B,2018ApJ...864..142D}, Ly$\alpha$ equivalent width distributions \citep{2018ApJ...856....2M,2019ApJ...878...12H,2019MNRAS.485.3947M}, stacked observations around heavily absorbed regions in the Ly$\alpha$ forest\citep{2024A&A...688L..26S,2024MNRAS.533L..49Z}, and the inferred reionization history from Planck polarization measurements assuming a parameterized tanh reionization history \citep{2021A&A...652C...4P}. All of our simulated reionization histories are broadly consistent with these current observational estimates.

The underlying matter density fields are generated using second-order Lagrangian perturbation theory (2LPT). For studies related to cosmic reionization, 2LPT should be sufficiently accurate. The simulation tracks $1024^3$ particles in a cubic volume with a comoving side length of 2 $h^{-1}$ Gpc.

\begin{table}[]
    \centering
    \caption{Reionization Simulation Parameters}
    \begin{tabular}{cccc}
    \hline
       \hline
       & $\bar{z}$   & $\alpha$ & $k_0 /\  {\rm hMpc^{-1}}$ \\
       \hline
      Fiducial &8 & 0.2&0.9\\
      \hline
      Late &7 &0.2 &0.9\\
      \hline
      Short & 8&0.564 &0.185\\
      \hline
    \end{tabular}
    \label{tab:simulation_para}
\end{table}

To generate an ionization field, we first use 2LPT to calculate the displacement of each particle in the simulations and move them to the midpoint of reionization $\bar{z}$. We then interpolate the particles onto a grid using triangular-shaped clouds (TSCs), and construct the matter density field $\delta_m(\vec{\mathbf{r}})$. Next, we compute the Fourier transform and apply the bias function in Equation~(\ref{eqn:bias}) to $\tilde{\delta}_m(\vec{\mathbf{k}})$. We then compute the inverse Fourier transform and generate the local redshift of cosmic reionization, $z_{\rm RE}(\vec{\mathbf{r}})$. Finally, the ionization field in each grid cell, at redshift $z_0$, is set to 1 if $z_{\rm RE}(\vec{\mathbf{r}})$ is greater than $z_0$, and 0 otherwise. In this work, we generate 30 independent realizations and average over them to reduce the cosmic variance in our signal estimates.

\subsection{21\,cm field}
\label{sec: 21cm_mock}
One can then use the ionization and density fields to generate 21\,cm brightness temperature fields according to \citep{1997ApJ...475..429M}:
\begin{equation}
    \delta T_{\rm 21cm}(\vec{\mathbf{r}},z) = T_0(z)[1+\delta_m(\vec{\mathbf{r}})][1-x_i(\vec{\mathbf{r}})]\,,
    \label{eqn:t21}
\end{equation}
where $T_0(z)$ is:
\begin{eqnarray}
\label{eqn:21cm_model}
    T_0(z) = 26 {\rm mK}&\left(\frac{T_S-T_\gamma}{T_S}\right)\left(\frac{\Omega_bh^2}{0.022}\right)\nonumber\\
    &\times\left[\left(\frac{0.143}{\Omega_mh^2}\right)\left(\frac{1+z}{10}\right)\right]^{\frac{1}{2}}.
\end{eqnarray}
Here, $T_S$ is the spin temperature and $T_\gamma$ is the CMB temperature. We assume that $T_S$ is globally coupled to the gas temperature, which has already been heated significantly above $T_\gamma$. Therefore, we can neglect the $(T_S-T_\gamma)/T_S$ factor in Equation~(\ref{eqn:21cm_model}). This assumption should be valid once the volume has been partially ionized.

The simulated 21\,cm fields are then used to generate 3D light cones across the entire reionization history. To obtain the features related to ``local'' redshift and study the redshift evolution of the cross-correlation, we need to truncate these 3D light cones into ``chunks'' with various midpoint redshift values $z_0$ and widths $\Delta z$, in preparation for subsequent data processing.

\subsection{kSZ field}
\label{sec: ksz_mock}
The kSZ effect is due to the inverse Compton scattering between CMB photons and free electrons with some peculiar velocity relative to the observer, $\mathbf{v}\cdot{\mathbf{\hat n}}$ where ${\mathbf{\hat n}}$ is the unit vector along the line of sight. The anisotropic CMB temperature fluctuation from the kSZ effect can be written as \citep{1972CoASP...4..173S}:
\begin{equation}
\label{eqn:cmb}
    \Delta T({\mathbf{\hat n}}) = - T_{\rm CMB}\int d\chi g(\chi){\mathbf{q}}\cdot{\mathbf{\hat n}}\,.
\end{equation}
Here $T_{\rm CMB}$ is the global average CMB temperature, $\chi$ is the comoving distance along the line of sight, $\mathbf{q}$ is the local electron momentum field with $\mathbf{q} = \mathbf{v}(1+\delta_m)(1+\delta_x)/c$, and $g(\chi)$ is the visibility function. We define $1+\delta_x = x_i/\langle x_i\rangle$, where $x_i$ is the local ionization fraction and $\langle x_i\rangle$ is the global average ionization fraction. The visibility function $g(\chi)$ describes the probability that a CMB photon scatters off a free electron between $\chi$ and $\chi-d\chi$. We can write this probability as:
\begin{equation}
\label{eqn:g(chi)}
    g(\chi) = \frac{\partial e^{-\tau(\chi)}}{\partial \chi}=e^{-\tau(\chi)}\sigma_Tn_{e,0}\langle x_i\rangle(1+z)^2\,,
\end{equation}
where $\tau(\chi)=\int\sigma_Tn_{e,0}\langle x_i\rangle(1+z)^2d\chi$ is the electron-scattering optical depth, $\sigma_T$ is the cross section for Thomson scattering, and $n_{e,0}$ is the mean electron number density at $z=0$. In principle, $\tau(\chi)$ should vary spatially, but we adopt the global average value for simplicity. This should be an excellent approximation since the $\tau$ fluctuations are small.  
We assume helium is singly-ionized along with hydrogen since helium is doubly-ionized only at lower redshifts \citep{2017ApJ...841...87L}. The extra free electrons produced when helium is doubly-ionized make only a small contribution to $\tau$ and can be safely neglected here. 

Next, we generate mock kSZ maps from the ionization fields obtained in Section~\ref{sec: simulation}.
We project the comoving coordinates in the simulation volume onto a fixed grid in angular coordinates ($\theta_x$, $\theta_y$). The matter density and velocity fields are obtained by interpolating between 2LPT simulation snapshots at two nearby redshift values. Finally, we integrate along the line of sight following Equations~(\ref{eqn:cmb}) and (\ref{eqn:g(chi)}). More details can be found in \citet{2022ApJ...928..162L}.

\subsection{Cross-Correlations}

Next, we consider the filtering steps which will be required with real data to mitigate foreground contamination and noise in each of the kSZ and 21\,cm fields. The filter we apply to the kSZ signal is the same as that in \citet{2022ApJ...928..162L}. This filter suppresses $\ell$ modes on scales where the patchy kSZ angular power spectrum is dominated by other contributions. Here we consider the lensed primary CMB spectrum, the late-time kSZ signal, residual foregrounds, and instrumental noise. Therefore, we can write the filter as:
\begin{equation}
    F^{\rm kSZ}(\ell) = \frac{C_\ell^{\rm kSZ,\ reion}}{C_\ell^{TT}+C_\ell^{\rm kSZ,\ reion}+C_\ell^{\rm kSZ,\ late}+N_\ell}\,,
\end{equation}
where $C_\ell^{\rm kSZ,\ reion}$ is the EoR component, $C_\ell^{\rm kSZ,\ late}$ is the post-reionization component, $C_\ell^{TT}$ is the lensed CMB primary spectrum, and $N_\ell$ combines both residual foregrounds and instrumental noise. Additionally, we account for the instrumental beam, i.e. for the finite angular resolution of the CMB measurements. We model the beam of the proposed experiment as a Gaussian beam and convolve the beam function $b(\ell)$ with the filter defined above. The final filtered kSZ fluctuation can be written as:
\begin{equation}
    \Delta T^{\rm kSZ}_f(\ell) = F^{\rm kSZ}(\ell)b(\ell)\Delta T^{\rm kSZ}(\ell)\equiv f^{\rm kSZ}(\ell)\Delta T^{\rm kSZ}(\ell)\,,
    \label{eqn:ksz_filter}
\end{equation}
where we define $f^{\rm kSZ}(\ell)\equiv F^{\rm kSZ}(\ell)b(\ell)$.

We use the standard internal linear combination (ILC) method to model the $N_\ell$ term. The ILC method constructs a weighted linear combination of intensity maps at different frequencies \citep{1992ApJ...391..466B,2003PhRvD..68l3523T,2004ApJ...612..633E,2009A&A...493..835D}. These weights should be configured to normalize the response to the desired blackbody spectrum and optimize the variance of the resulting map. We use the publicly available {\tt orphics\footnote{\url{https://github.com/msyriac/orphics}}} code \citep{2021PhRvD.104h3529H}, which applies the ILC method in harmonic space, to estimate the combined contribution of residual CIB, radio sources, tSZ, and instrumental noise. 

As mentioned above, 21\,cm observations suffer from foreground contamination, which can be 4-5 orders of magnitude larger than the cosmological signal from the EoR \citep{2012ApJ...756..165P,2017PASP..129d5001D}. The foregrounds are expected to roughly follow a power-law in frequency. Therefore, in principle, these bright foregrounds should intrinsically occupy only low-$k_\parallel$ modes in Fourier space. However, in practice, the foregrounds leak into high-$k_\parallel$ modes due to the chromatic response of the interferometer. The leakage spans a ``wedge'' shape in ($k_\perp,\ k_\parallel$) space, and the contamination can be characterized as a linear relationship between $k_\perp$ and $k_\parallel$.
We describe the wedge by its slope, $m(z)$, and express the relationship as $k_\parallel = m(z) k_\perp$. In a ``horizon wedge'' scenario, the slope should be:
\begin{equation}
    \label{eqn: wedge}
    m(z) = \frac{H(z)D_c}{c(1+z)},
\end{equation}
where $H(z)$ is the Hubble parameter, $D_c$ is the line-of-sight comoving distance, and $c$ is the speed of light. All modes with $k_\parallel \gtrsim m(z) k_\perp$ are assumed to be contaminated by foregrounds.
During the EoR, $m$ is around 3. In this work, we also discuss scenarios with smaller $m$, assuming that foregrounds can be partially removed inside the horizon wedge.

To avoid the contamination of such wedge-like foregrounds in Fourier space and to reduce the impact of instrumental noise, we apply another filter to the 21\,cm fields. In general, we consider two scenarios here. One is an optimistic scenario where we assume that the chromatic response of the interferometer can be calibrated and that foregrounds only corrupt low-$k_\parallel$ modes. In this scenario, we can write the foreground-avoiding component of the filter as:
\begin{equation}
    f^{\rm 21cm}_{\rm fore,\,opti}(k_\perp,\ k_\parallel) = 
    \begin{cases}
    1 & k_\parallel> k_{\parallel,0}\ h{\rm Mpc^{-1}},\\
    0 & \rm{Otherwise}\,.
    \end{cases}
\end{equation}
In an alternative more realistic scenario, we assume that the foreground leakage is restricted to lie within a wedge in Fourier space. In this scenario,  the foreground component of the filter is:
\begin{equation}
    \label{eqn: wedge_filter}
    f^{\rm 21cm}_{\rm fore,\,real}(k_\perp,\ k_\parallel) = 
    \begin{cases}
    1 & k_\parallel> k_\perp m,\\
    0 & \rm{Otherwise}\,.
    \end{cases}
\end{equation}
We regard $k_{\parallel,0}$ as the ``optimistic'' scenario and use the wedge slope $m$ to parametrize foreground contamination in the realistic scenario. We will discuss the filter size dependence in Section~\ref{sec: results}.

For the instrumental noise component, we adopt a Wiener filter determined by the ratio between the cosmological signal power spectrum and the noise power spectrum in cylindrical ($k_\perp,\ k_\parallel$) space:
\begin{equation}
    f^{\rm 21cm}_{\rm noise}(k_\perp,\ k_\parallel) = \frac{P^{\rm 21cm,\ reion}(k_\perp,\ k_\parallel)}{P^{\rm 21cm,\ reion}(k_\perp,\ k_\parallel)+P^{\rm 21cm,\ noise}(k_\perp)}
\end{equation}

The noise term depends on the observational strategies adopted and on the instrumental configuration. In this work, we consider two interferometric arrays, HERA and SKA1-Low, which are expected to detect the EoR 21\,cm signal in the current decade. The proposed observational strategies and technical details of HERA have been fully discussed, determined, and implemented. We model the noise power spectrum of HERA as follows \citep{2014ApJ...788..106P}:
\begin{equation}
    P^{\rm 21cm,\ HERA}_{\textbf{k}} = \frac{T^2_{\rm sys}\Omega_p^2(\nu)X^2(\nu)Y(\nu)}{\Omega_{pp}(\nu)t_{\rm int}N_{\rm pol}N_{\rm bl}(u)}\,,
    \label{eqn:hera-noise}
\end{equation}
where $T_{\rm sys}$ is the system temperature of the interferometer, $\Omega_p$ is the integral of the primary beam of the antenna, $\Omega_{pp}$ is the integral of the square of the primary beam, $t_{\rm int}$ is the observation time, $N_{\rm pol}$ is the number of independent polarizations, and $N_{\rm bl}$ is the baseline number counts. The quantities $X(\nu)$ and $Y(\nu)$ are factors that account for the cosmological units in the plane of the sky and along the line of sight respectively:
\begin{eqnarray}
    X(\nu) = \frac{dr_{\perp}}{d\ell} = D_M(\nu),\\
    Y(\nu) = \frac{dr_{||}}{d\nu} = \frac{c(1+z)}{H(z)\nu},
\end{eqnarray}
where $D_M$ is the transverse co-moving distance. For a flat Universe, $D_M = D_c$.

Work on SKA1-Low is ongoing and, as such, the construction plan and observing strategies are not fully decided yet. 
However, as a starting point for the purposes of generating forecasts, we model the noise power spectrum of SKA1-Low as follows \citep{2006ApJ...653..815M,2013PhRvD..88h1303M}:
\begin{equation}
    P^{\rm 21cm,\ SKA}_{\textbf{k}} = \frac{\lambda^2T^2_{\rm sys}X^2(\nu)Y(\nu)}{t_{\rm int}N_{\rm bl}(u)A_e}\,,
    \label{eqn:ska-noise}
\end{equation}
where $A_e$ is the effective collecting area and $\lambda$ is the wavelength of observation. We do not correct the beam area with the $\Omega^2_p/\Omega_{pp}$ in Equation~(\ref{eqn:ska-noise}) because the details of SKA1-Low are not yet well-determined.
\begin{figure}
    \centering
    \includegraphics[width=0.9\columnwidth]{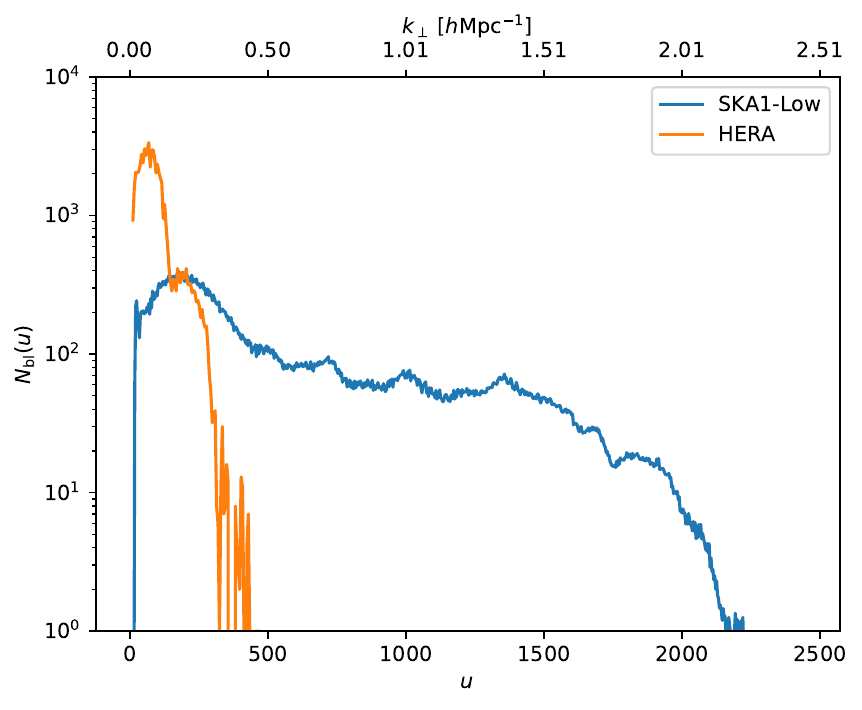}
    \caption{Baseline number counts of HERA and SKA1-Low as a function of $u$. We also show the corresponding $k_\perp$ at $z = 8.00$ on the upper x-axis.}
    \label{fig:baseline_counts}
\end{figure}

\begin{figure*}
    \centering
    \includegraphics[width=0.9\columnwidth]{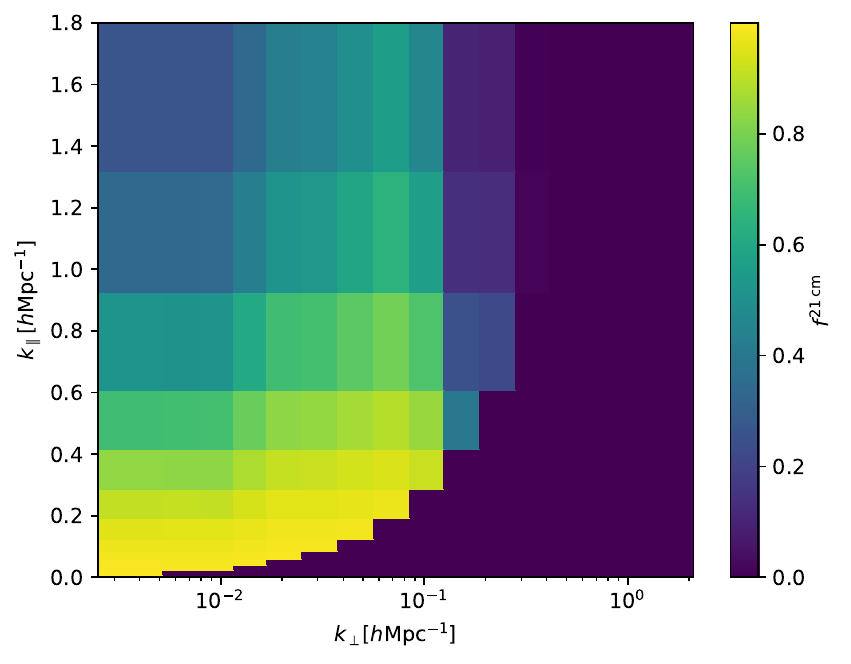}
    \includegraphics[width=0.9\columnwidth]{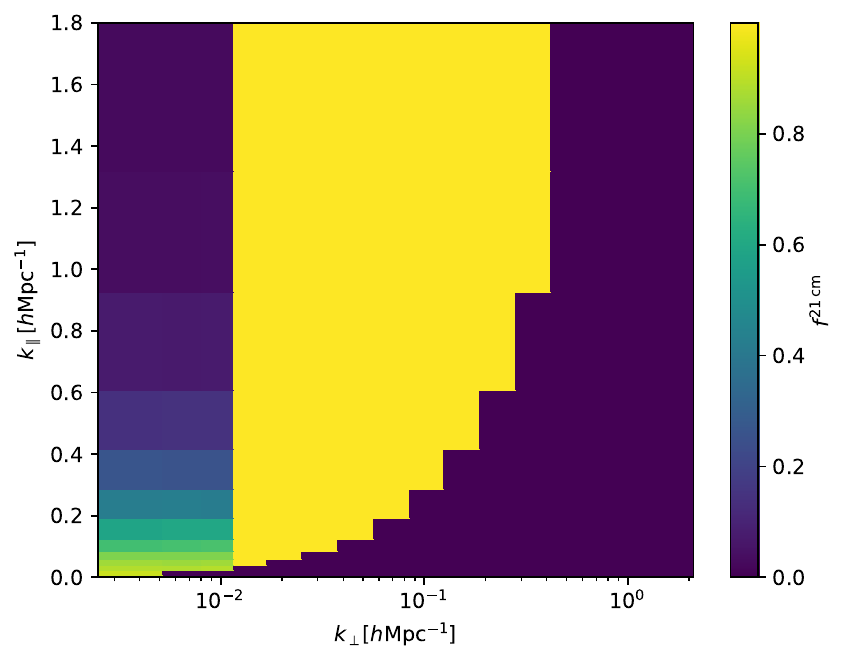}
    \caption{The 21\,cm filter with $m=3$ and $\Delta z=2.0$, at $z_0=8.00$, assuming HERA (left) and SKA1-Low (right) noise respectively. The foreground components filter out the wedge region in both cases. In the case of HERA, the noise component filter suppresses the large scale modes while in the case of SKA1-Low it suppresses the small scale modes.}
    \label{fig:filters}
\end{figure*}

\begin{figure*}
    \centering
    \includegraphics[width=0.9\columnwidth]{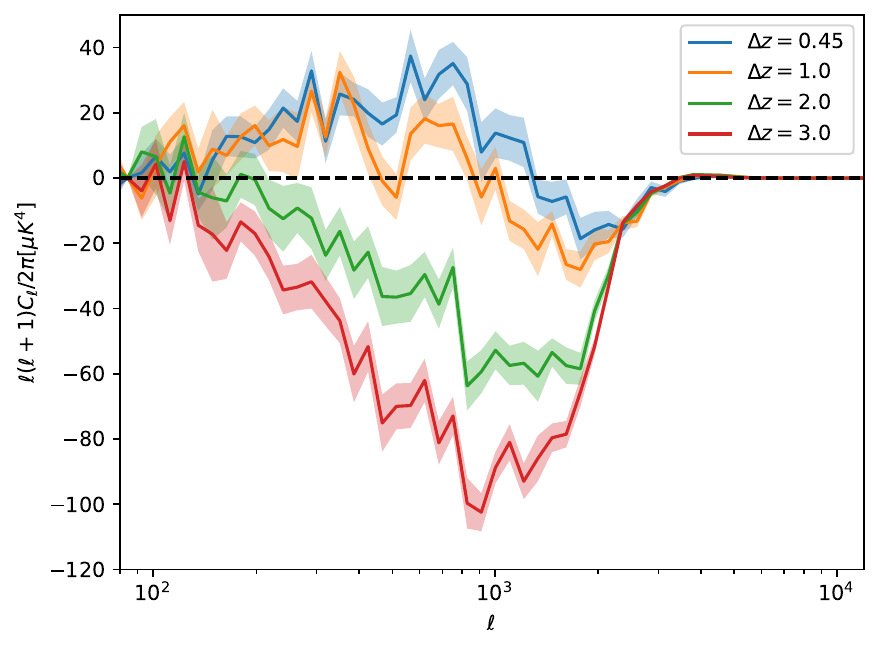}
    \includegraphics[width=0.9\columnwidth]{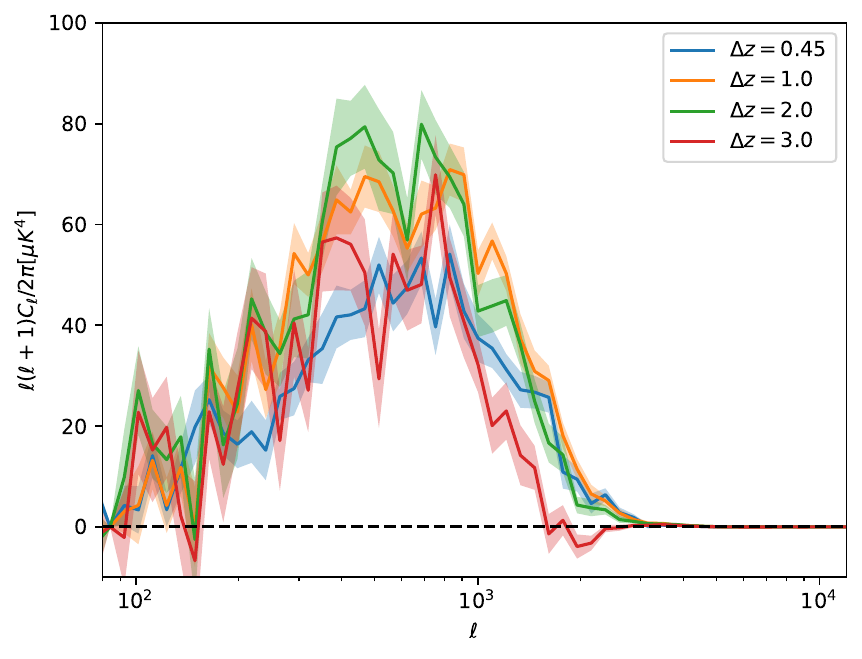}
    
    \includegraphics[width=0.9\columnwidth]{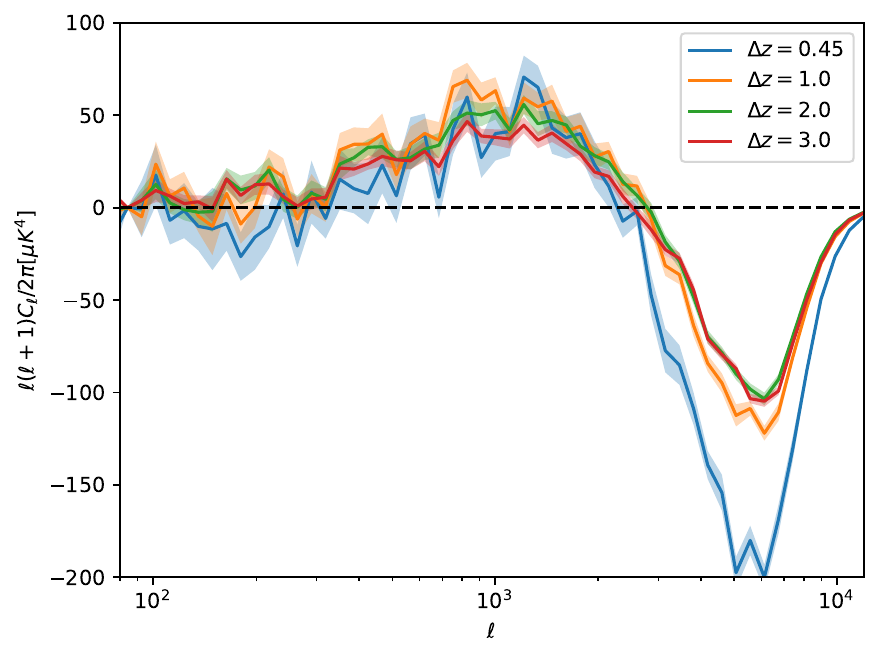}
    \includegraphics[width=0.9\columnwidth]{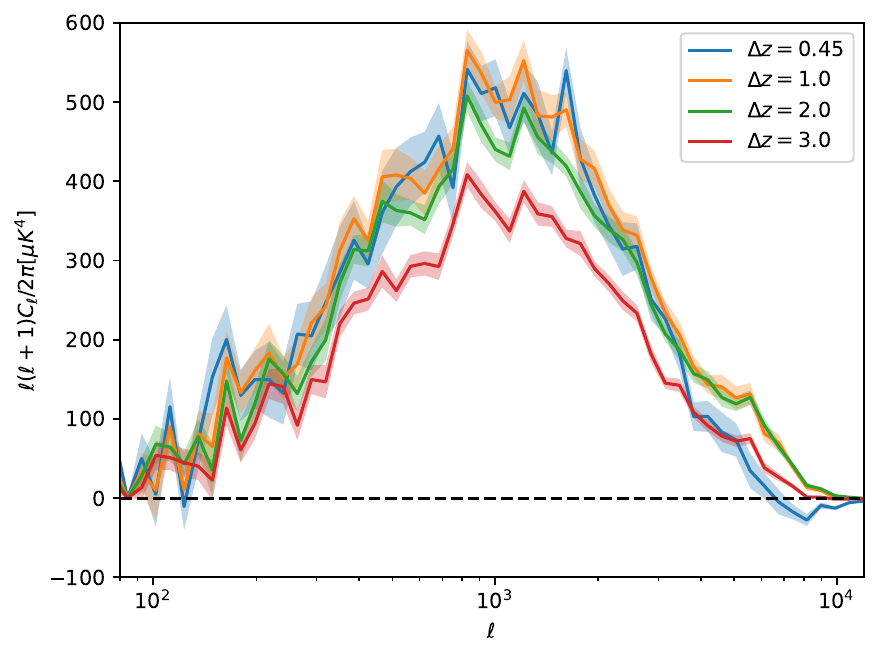}
    \caption{The $\rm kSZ^2$-$\rm 21cm^2$ cross-power spectrum $C_\ell^{\rm kSZ^2\times\rm 21cm^2}$ as a function of $\ell$ at different redshifts, assuming an optimistic foreground filter with $k_{||,0}=0.01$. The cross-power spectrum is estimated at a central redshift of $z_0=8.00$ (left) and $9.00$ (right) respectively, corresponding to volume-averaged ionization fractions of $\bar{x}_{\rm HI} = 0.43$, and $0.20$. We apply the HERA noise filter in the upper panels and SKA1-Low noise filter in the lower panels. We also vary the width ($\Delta z$) at each redshift (see labels in the legend). The shaded bands show the standard error over 30 independent simulation realizations.}
    \label{fig:chunksize}
\end{figure*}

In both cases, we assume a frequency-dependent system temperature of $T_{\rm sys} = 237+1.6(\nu/300{\rm MHz})^{-5.23} {\rm K}$\footnote{This fitting result is obtained by assuming a template of $T_{\rm sys}(\nu)=a+b({\nu/300{\rm MHz}})^{c}$. The spectral index fit is much larger than a synchrotron-like power law index, i.e. -2.6. 
This is because the major contribution is from the constant component and the uncertainty in the fitted spectral index is very large.
}, which is determined based on \citet{2021ApJS..255...26T}. In the case of HERA, we assume an observing time of $t_{\rm int} = 200$ hours and $N_{\rm pol} = 2$. We also use the calculations of $N_{\rm bl}$ and the ratio $\Omega^2_p/\Omega_{pp}$ in \citet{2023ApJ...944...59L}. In the case of SKA1-Low, we use {\tt Tools21cm\footnote{\url{https://github.com/sambit-giri/tools21cm}}} \citep{2020JOSS....5.2363G} to estimate the baseline number counts. We assume a collecting area of $A_e = 962 {\rm m^2}$ and an observing time of $t_{\rm int} = 1000$ hours. In Figure~\ref{fig:baseline_counts}, we plot the baseline counts of HERA and SKA1-Low under the configuration described above, as a function of $u$ and $k_\perp$. For HERA, most baselines measure large-scale modes while for SKA1-Low, smaller-scale modes will be probed.

The final filtered 21\,cm fluctuation field can be written as:
\begin{equation}
    \Delta T^{\rm 21cm}_f(\mathbf{k}) = f^{\rm 21cm}_{\rm fore}f^{\rm 21cm}_{\rm noise}\Delta T^{\rm 21cm}(\mathbf{k})\equiv f^{\rm 21cm}\Delta T^{\rm 21cm}(\mathbf{k})\,,
    \label{eqn:t21_filter}
\end{equation}
where we define $f^{\rm 21cm}(\textbf{k})\equiv f^{\rm 21cm}_{\rm fore}f^{\rm 21cm}_{\rm noise}$. For the purpose of illustration, we plot two different filters with $m=3$ and $\Delta z=2.0$, at $z_0=8.00$ in Figure~\ref{fig:filters}, assuming HERA (left) and SKA1-Low (right) noise, respectively. The foreground components filter out the wedge region in both cases. In the case of HERA, the noise component filter suppresses the large-scale modes while in the case of SKA1-Low it suppresses the small-scale modes.

As a summary, we compute the ${\rm kSZ^2}$--${\rm 21cm^2}$ cross-power spectrum from a simulated two-dimensional kSZ map and a three-dimensional 21\,cm light cone using the following procedure:
\begin{enumerate}
\item We extract a given range of redshifts (a ``redshift chunk'') from the full 21\,cm light cone, using the 21\,cm redshift window discussed in Section~\ref{sec: simulation}.
\item We Fourier transform each of the two-dimensional kSZ map and the three-dimensional 21\,cm chunks. 
\item We apply Fourier space filters using Equation~(\ref{eqn:ksz_filter}) for the kSZ field and Equation~(\ref{eqn:t21_filter}) for the 21\,cm fields. We then inverse Fourier transform.
\item We square the resulting two-dimensional kSZ map and the three-dimensional 21\,cm fields in configuration space.
\item We project each three-dimensional 21 cm redshift chunk into a two-dimensional map using the same top-hat redshift window, of width $\Delta z$, used to define each redshift chunk. We then Fourier transform each two-dimensional kSZ and 21\,cm map.
\item Finally, we compute the cross-power spectrum between the two filtered and then squared fields.
\end{enumerate}

Note that the order of operations is important here. If one performs the projection before squaring the 21\,cm field, the residual radial modes that survived the filtering step will be lost.

\section{Results} \label{sec: results}
Next, we show the simulated ${\rm kSZ}^2$-${\rm 21cm}^2$ cross-power spectrum signals, and illustrate their dependence on redshift, and the precise filters adopted. We also explore how the signals evolve as the reionization process proceeds. In this section, we apply the SO noise filter and the HERA noise filter to the kSZ and 21\,cm fields, respectively. As discussed further in Section~\ref{sec: detectability}, we later vary these filters to generate forecasts for different survey combinations.

\subsection{Dependence on chunk size}\label{subsec: chunk}
We first consider the dependence on the ``chunk size'', i.e. on the redshift width of each chunk, $\Delta z$. The $\Delta z$ dependence here should reflect how much the squared 21\,cm field evolves across the redshift chunk of interest. 

To present the dependence on chunk size, we plot the angular cross-power spectrum between the filtered ${\rm kSZ^2}$ and ${\rm{21cm^2}}$ signals, $C_\ell^{\rm kSZ^2\times21cm^2}$, in the fiducial reionization scenario as a function of $\ell$ in Figure~\ref{fig:chunksize}. Here we employ the foreground filter in the optimistic scenario with $k_{\parallel,0}=0.01$. 
We apply the HERA noise filter in the upper panels and the SKA1-Low noise filter in the lower panels. We vary the midpoint redshift $z_0$ and the width $\Delta_z$. $C_\ell^{\rm kSZ^2\times21cm^2}$ is computed at the central redshift $z_0 = 8.0$ (left) and $z_0 = 9.0$ (right) respectively. In each panel, we also vary the redshift width $\Delta z$ as shown by the legend. At high redshift ($z_0 = 9.0$), where the squared 21\,cm field evolves slowly, the chunk size dependence is weak. As the chunk size increases, the shape of $C_\ell^{\rm kSZ^2\times21cm^2}$ as a function of $\ell$ remains nearly the same, and the amplitude barely decreases unless the chunk size is too big. 
However, at lower redshifts, we find significant changes in the amplitude of $C_\ell^{\rm kSZ^2\times21cm^2}$ as the chunk size increases. 
We also notice that in the case of HERA, the signal vanishes beyond $\ell\sim 3000$. This is because HERA lacks long baselines, capable of probing small-scale modes.
Therefore the HERA noise filter strongly suppresses these $\ell>3000$ modes. Another interesting feature is that the sign of $C_\ell^{\rm kSZ^2\times21cm^2}$ flips for some $\ell$-values at $z_0 = 8.00$, while at $z_0=9.0$, the cross-spectrum is always positive. 
The scale, $\ell$, at which this statistic changes sign is likely related to the characteristic size of the ionized bubbles at the corresponding stage of reionization. 
The reason we suspect this connection between the sign-flip scale and the characteristic bubble size is that they have similar dependencies on chunk size. In the case of SKA, even though the typical bubble size should evolve with redshift, the effective bubble size in a chunk with a fixed central redshift should be similar. In the case of HERA, as already mentioned above, modes above $\ell\sim3000$ are removed by the noise filter. Small bubbles at the high redshift end in a chunk can be smoothed out. In this case, once the chunk size increases, the effective bubble size should also increase. As Figure \ref{fig:chunksize} shows, the sign-flip scales in both cases also have similar dependencies. This suggests that the sign-flip feature is indeed related to the typical bubble size.
\begin{figure*}
    \flushleft
    \qquad\qquad
    \includegraphics[width=0.6\columnwidth]{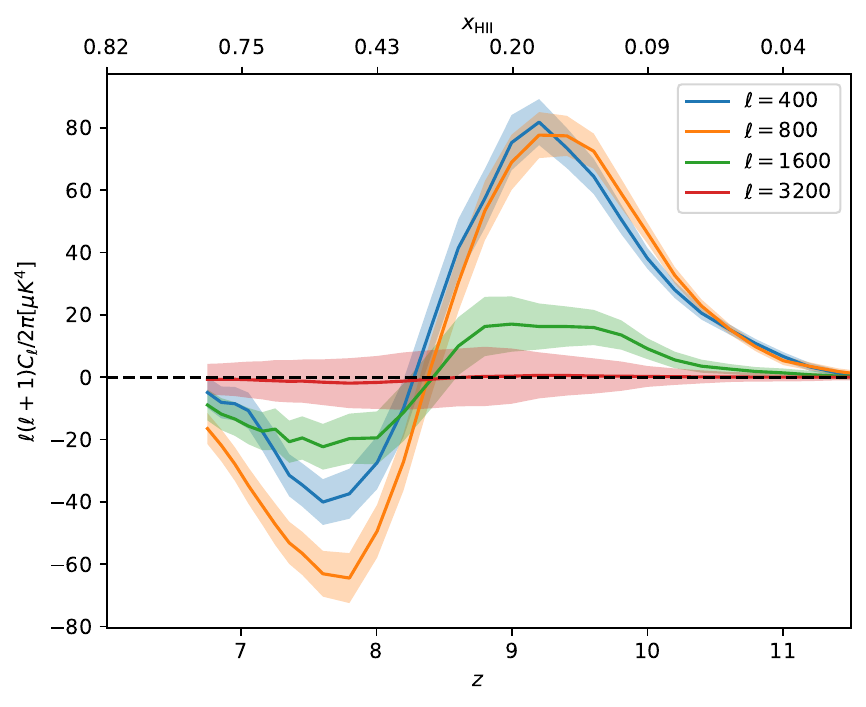}
    \includegraphics[width=0.6\columnwidth]{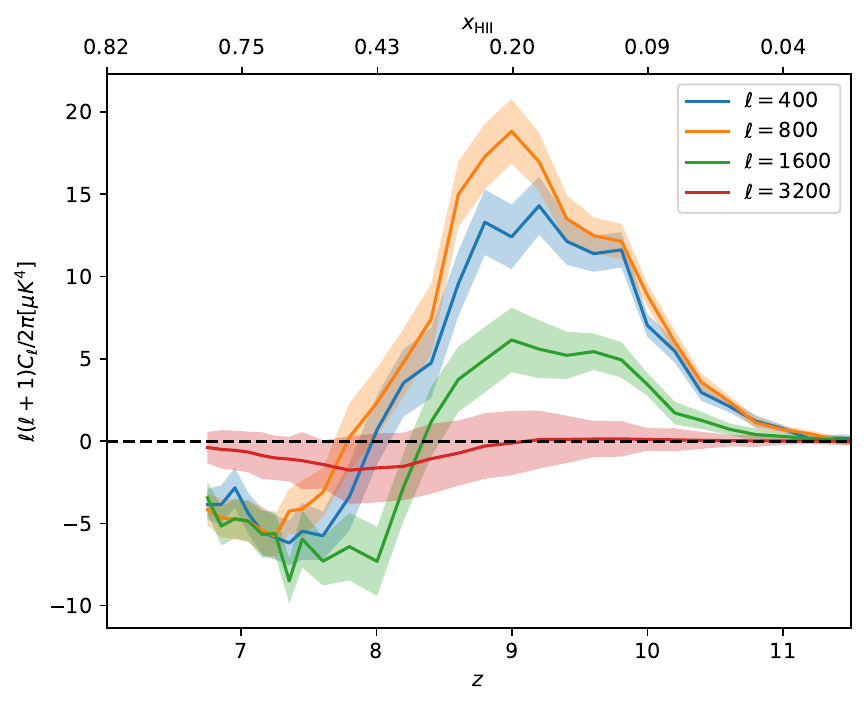}
    
    \centering
    \includegraphics[width=0.6\columnwidth]{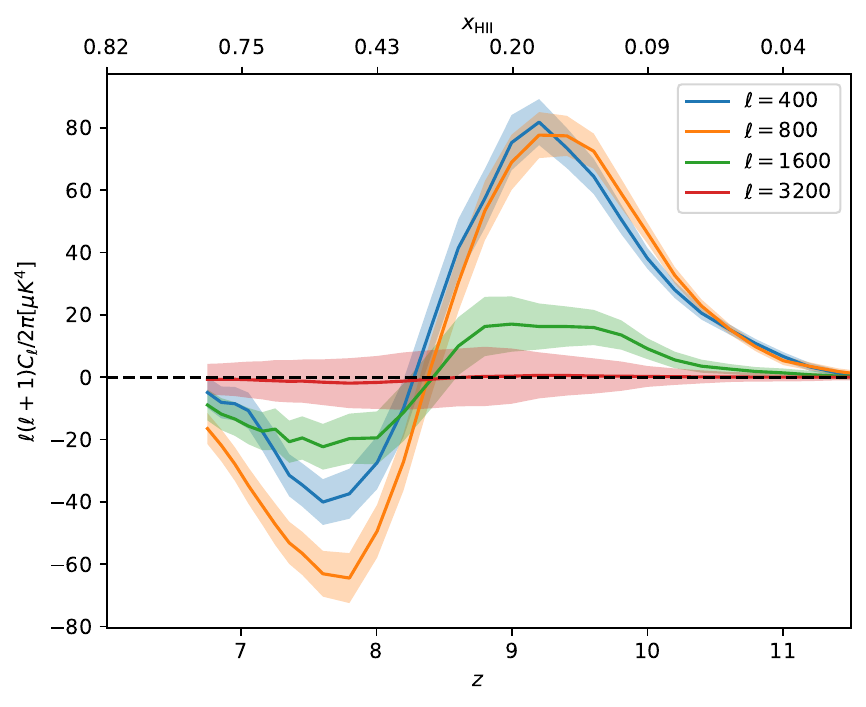}
    \includegraphics[width=0.6\columnwidth]{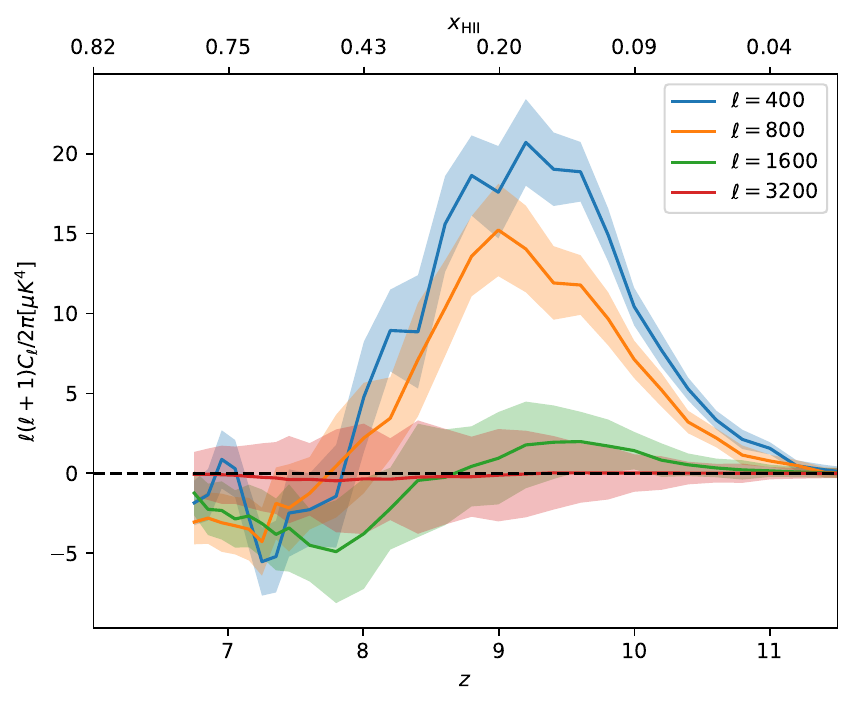}
    \includegraphics[width=0.6\columnwidth]{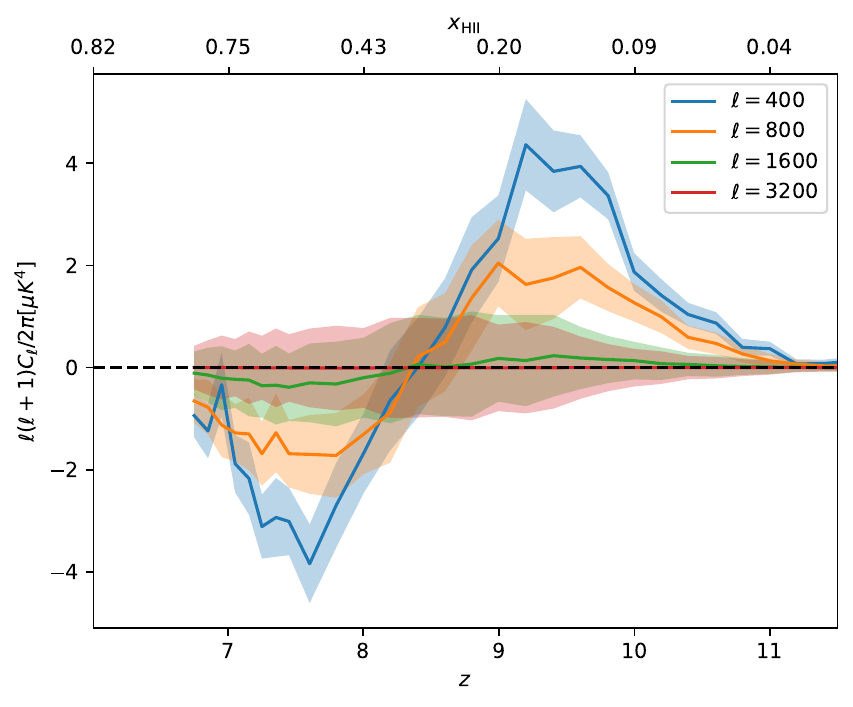}
    \caption{$C_\ell^{\rm kSZ^2\times\rm 21cm^2}$ as a function of $z_0$. We fix the chunk size to $\Delta z=2.0$ and employ filters based on HERA and SO specifications. We apply the optimistic foreground filters with (from left to right) $k_{\parallel,0}=0.01$ and $0.10$ in the upper panels and the realistic foreground filters with (from left to right) $m=0.1$, $1.0$, and $3.0$ in the lower panels, respectively. The corresponding volume-averaged ionization fractions, $\bar{x}_{\rm HII}$, are plotted on the upper x-axis. The shaded regions show the standard errors as in Figure~\ref{fig:chunksize}.}
    \label{fig:filtersize}
\end{figure*}

\subsection{Dependence on foreground filter size}
In this section, we estimate the impact of the foreground filter size on the magnitude of the cross-correlation signal. In principle, the scales employed in the foreground filter represent a trade-off. A larger-scale filter removes more foreground power, but also more cosmological signal power. On the other hand, a smaller-scale filter preserves more cosmological signal power but also more residual foreground power. For the purposes of this analysis, however, we do not include any residual foregrounds in our simulations outside of the filtered modes. Implicitly, this assumes that the foregrounds have been avoided entirely (e.g., by using a wedge-like filter) or that instrumental calibration steps have successfully reduced foregrounds to a level well below that of the signal. In the following analysis, we treat the different 21\,cm filters as different cases for how much useful information can be extracted from the 21\,cm measurements. 
In this way, we believe that the foreground wedge with a slope of $m = 3.0$ is a pessimistic ``worst-case'' for the prospective signal, and other filters represent improvements to this case.

Therefore, we vary the filter size in both scenarios and plot the corresponding cross-power spectrum in the fiducial reionization scenario, as a function of redshift in Figure~\ref{fig:filtersize}. We fix the chunk size to $\Delta z=2.0$ in all cases. We employ the optimistic foreground filter in the upper panels and the realistic foreground filter in the lower panels. In the optimistic scenario, from left to right, we show the results with $k_{\parallel,0}=0.01$ and $0.10$, respectively. In the realistic scenario, from left to right, we show the results with $m=0.1$, $1.0$, and $3.0$, respectively. 

We find similar redshift evolution trends in all cases and at all scales. As cosmic reionization proceeds, $C_\ell^{\rm kSZ^2\times21cm^2}$ first increases and reaches a maximum around $z=9.0$. The signal then decreases until it reaches a minimum around $z=7.6$. Finally, it increases to zero and vanishes at the end of reionization. The amplitude of $C_\ell^{\rm kSZ^2\times21cm^2}$ also has some scale dependence. The amplitude is greater at $\ell\sim400$ and $800$, falls to a lower level at $\ell\sim1600$, and almost disappears by $\ell\sim3200$. 
We also find that during later stages of reionization, the sign-flip occurs at larger scales. This is consistent with the results in Section~\ref{subsec: chunk}. 
The sign-flip is related to the typical sizes of the ionized bubbles, which grow larger as reionization proceeds, and so the sign flips occur at lower $\ell$ during the late stages of reionization.

Figure~\ref{fig:filtersize} also shows that the amplitude of the cross-power spectrum decreases for more severe foreground avoidance filtering (i.e. as $m$ increases, see Equations~(\ref{eqn: wedge}) and~(\ref{eqn: wedge_filter})). The signal also becomes a less smooth function of redshift. The reduction in the amplitude is because the removed Fourier modes intrinsically have a larger amplitude, and so without their contribution the resulting signal strength is smaller. We also break the internal statistical covariance relationship of the 21\,cm field in Fourier space more severely. Put differently, the remaining modes show weaker correlations between the 21\,cm and kSZ fields. Therefore, we lose smoothness once we project the 3D 21\,cm field onto a 2D map and correlate it with the kSZ map. Nevertheless, the fact that the kSZ field is inherently two-dimensional requires this projection as part of the cross-spectrum data analysis procedure.

\begin{figure*}
    \centering
    \includegraphics[width=0.9\columnwidth]{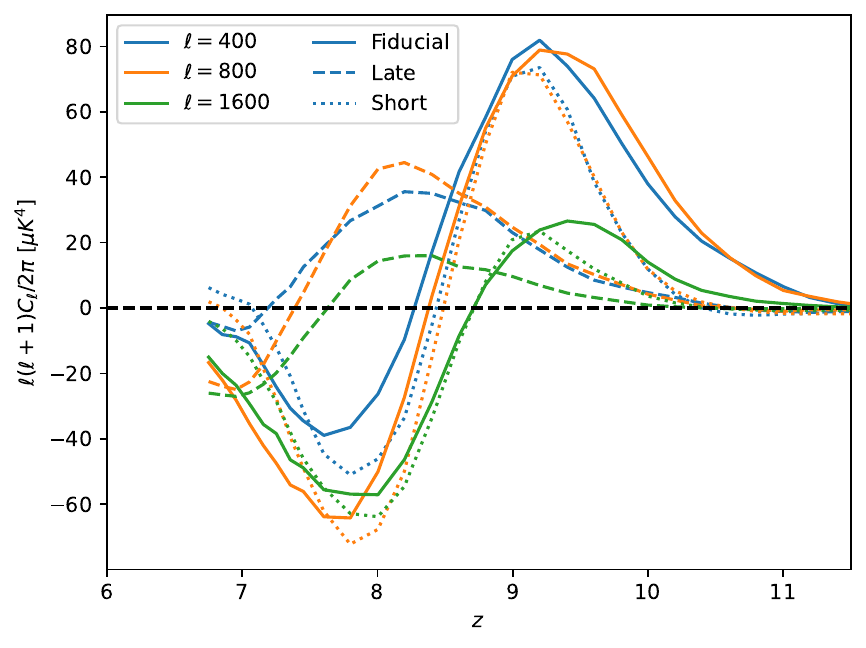}
    \includegraphics[width=0.9\columnwidth]{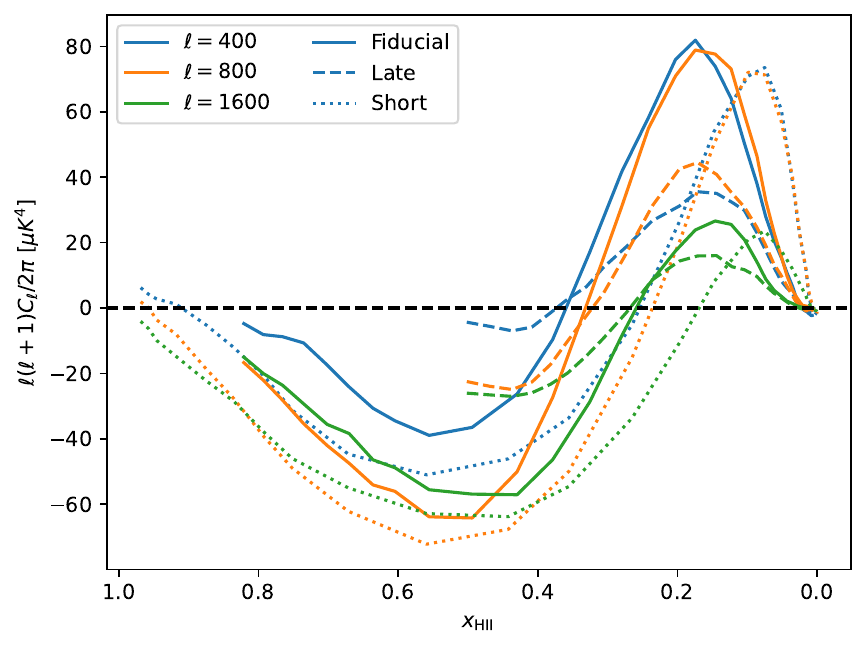}
    
    \includegraphics[width=0.9\columnwidth]{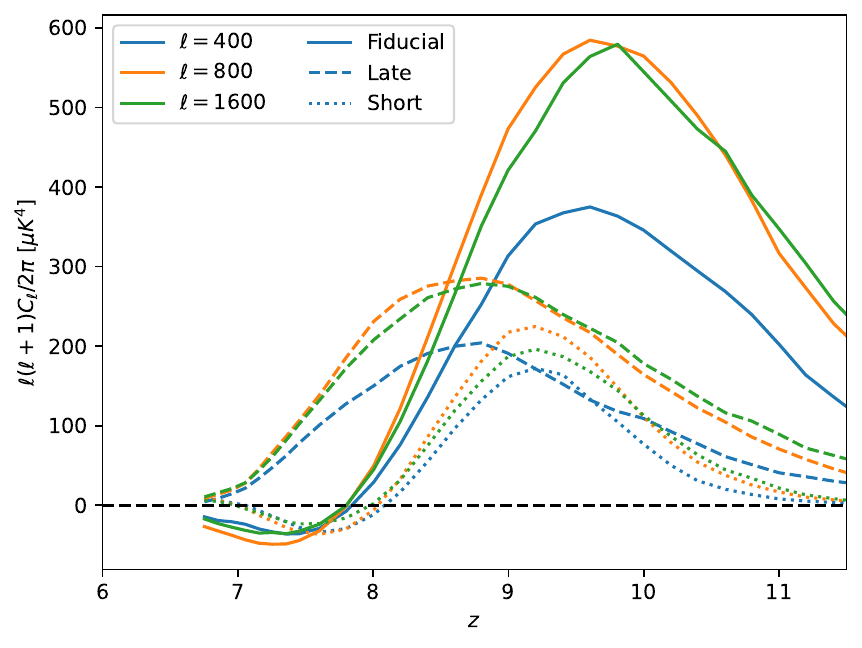}
    \includegraphics[width=0.9\columnwidth]{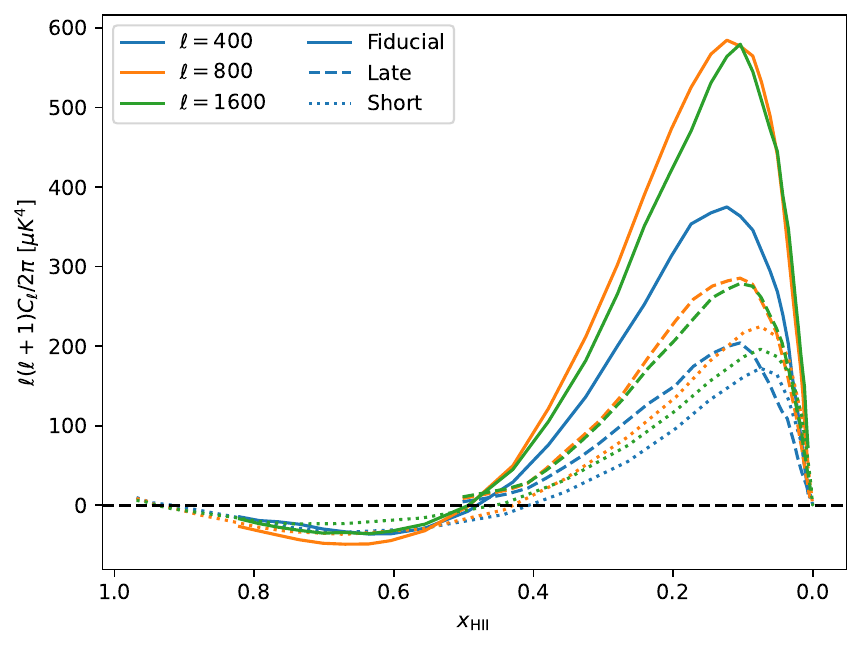}
    \caption{$C_\ell^{\rm kSZ^2\times\rm 21cm^2}$ as a function of $z_0$ (left) and $\bar{x}_{\rm HII}$ (right), respectively. The late scenario does not extend to the end of reionization because the stored mock lightcones end at $z=6$. The ionization fraction at the center of the lowest redshift chunk is around $\bar{x}_{\rm HII} \sim 0.5$ in the late scenario. We use the optimistic filter with $k_{||,0}=0.01$ and adopt the different reionization histories shown in Figure~\ref{fig:hist}. We apply the HERA noise filter in the upper panels and the SKA1 noise filter in the lower panels, respectively.  We find that in the case of SKA1, the signals share similar evolution with respect to $\bar{x}_{\rm HII}$. Their amplitudes are confirmed to be proportional to $(C_\ell^{\rm kSZ^2\times\rm kSZ^2})^{\frac{1}{2}}$. In the case of HERA, the filter breaks the reionization history dependence and amplitude relation due to its strong scale dependence.}
    \label{fig:historydependence}
\end{figure*}

\subsection{Dependence on ionization history} \label{subsec: hist}
In this subsection, we explore how the cross-power spectrum changes under variations in the reionization history model assumed. Specifically, we vary the parameters in our simulations and produce two additional reionization histories, denoted the ``short'' and ``late'' reionization models (see Figure~\ref{fig:hist} and Table~\ref{tab:simulation_para} for more details). 

In Figure~\ref{fig:historydependence}, we compare the features of $\rm kSZ^2$-$\rm 21cm^2$ cross-power spectrum in the three different reionization models. 
We apply the HERA noise filer in the upper panels and the SKA1-Low noise filter in the bottom panels. In all cases, we  use the optimistic filter with $k_{\parallel,0}=0.01$ and fix the chunk size to $\Delta z=2.0$. In the left panels, we plot the signal as a function of redshift. In the case of HERA, the signal in the late scenario is the smallest, while the signals in the fiducial and the short scenario have similar amplitude and redshift evolution. In the case of SKA1-Low, the signal in the short scenario is the smallest while the signal in the fiducial scenario is the largest. Without filtering, the amplitude of the cross-power spectrum should depend on the amplitude of the squared kSZ signals since the amplitudes of the squared 21\,cm signals are similar in all scenarios. Strictly speaking, it should be proportional to $(C_\ell^{\rm kSZ^2\times\rm kSZ^2})^{\frac{1}{2}}$. We evaluate the ratios of this quantity between three reionization scenarios and find that $(C^{\rm Late}/C^{\rm Fiducial})^{\frac{1}{2}}\sim 0.5$ and $(C^{\rm Short}/C^{\rm Fiducial})^{\frac{1}{2}}\sim 0.3$. These ratios are consistent with the ratios between $C_\ell^{\rm kSZ^2\times\rm 21cm^2}$ in the three reionization scenarios in the case of SKA1-Low, whose filter is near unity at all scales. In the case of HERA, the noise filter breaks this relation due to its strong scale dependence.

In the right panel of Figure~\ref{fig:historydependence}, we also plot the cross-power spectrum as a function of the volume-averaged ionization fraction. The signals in the three different models share a similar evolution with respect to the average ionization fraction in the case of SKA1. 
They peak around $x_{\rm HII} \sim 0.15$ and reach their lowest points around the midpoint of reionization. However, we find that the HERA filter results in some differences between the three reionization models. The short scenario peaks earlier around $x_{\rm HII} \sim 0.10$ while the other two peak around $x_{\rm HII} \sim 0.20$.

\section{DETECTABILITY}\label{sec: detectability}
In this section, we estimate the detectability of this signal in future surveys. As mentioned above, we propose combining the 21\,cm fluctuations measured by HERA or SKA1-Low with the kSZ signal from SO, CMB-S4, or CMB-HD. First, we estimate the total S/N of one experimental combination in a given redshift bin $z_0$ after summing (in quadrature) over all $\ell$ modes. Using our results in Section~\ref{sec: results} as the ``true'' value of this cross-power spectrum signal, the total S/N of one measurement can be written as \citep{2016PhRvD..94l3526F,2022ApJ...928..162L}:
\begin{equation}
    \left(\frac{\rm S}{\rm N}\right)^2_{z_0} = f_{\rm sky}\sum_\ell\frac{(2\ell+1)(C_\ell^{\rm kSZ^2\times21cm^2})^2}{C_\ell^{T^2T^2,f}C_\ell^{{\rm{21cm^221cm^2}},f}+(C_\ell^{\rm kSZ^2\times21cm^2})^2}\,,
\end{equation}
where $f_{\rm sky}$ is the fraction of overlapping sky surveyed by both experiments, $C_\ell^{T^2T^2,f}$ is the total power spectrum of the filtered ${\rm CMB^2}$ map, and $C_\ell^{{\rm{21cm^221cm^2}},f}$ is the angular power spectrum of the filtered 21\,cm projected map including instrumental noises. We apply the same approach in \citet{2016PhRvD..94l3526F} and \citet{2022ApJ...928..162L} to estimate $C_\ell^{T^2T^2,f}$ in the Gaussian approximation:
\begin{equation}\label{kszgaussian}
    C_{\ell}^{T^2T^2,f} \approx 2\int \frac{d^2\boldsymbol{L}}{(2\pi)^2} C_{\boldsymbol{L}}^{TT,f}C^{TT,f}_{|\boldsymbol{L-\ell}|}\,,
\end{equation}
\begin{figure*}
    \centering
    \includegraphics[width=0.9\columnwidth]{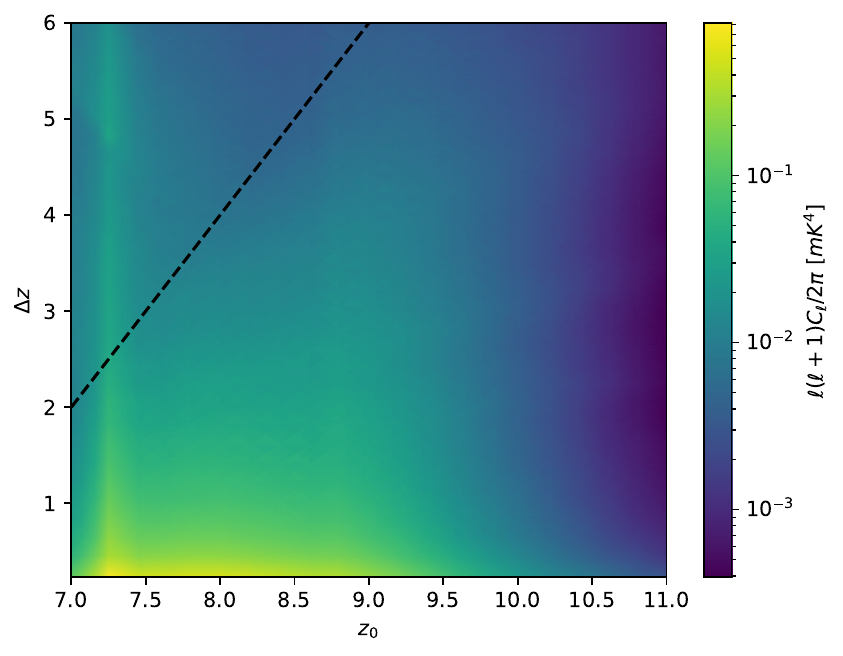}
    \includegraphics[width=0.9\columnwidth]{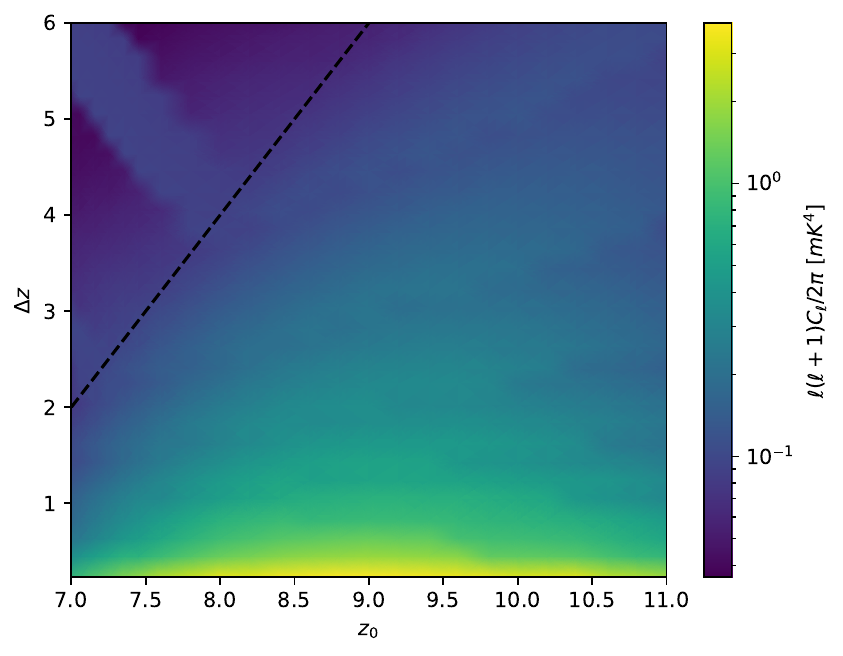}
    
    \includegraphics[width=0.9\columnwidth]{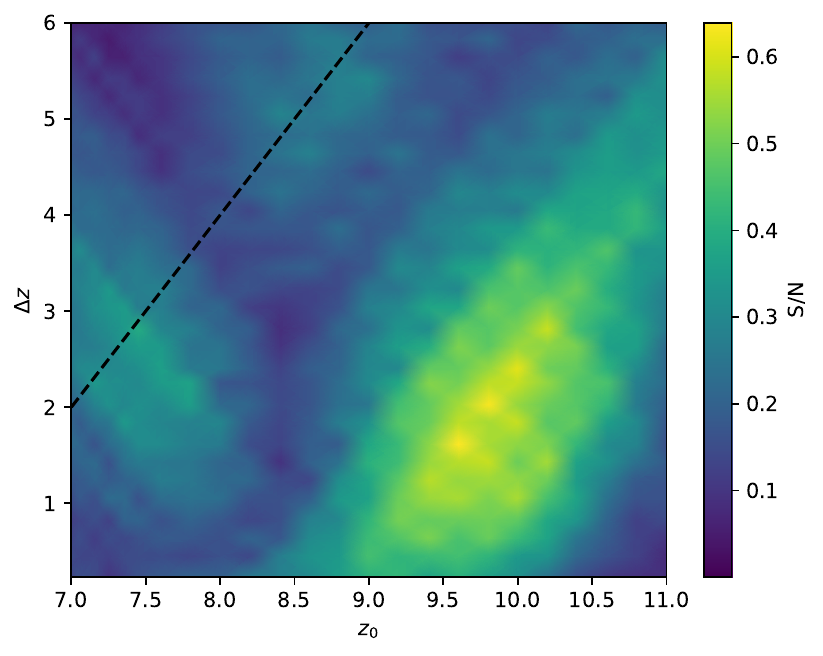}
    \includegraphics[width=0.9 \columnwidth]{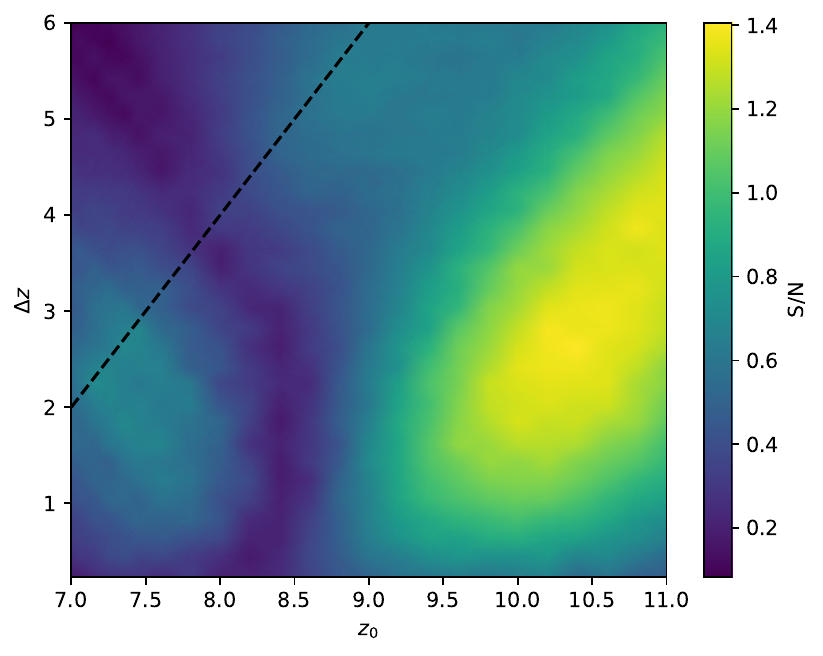}
    \caption{$C_\ell^{{\rm{21cm^221cm^2}},f}$ (upper), when $\ell=1600$, and S/N (bottom), as a function of $z_0$ and $\Delta z$. We forecast the expected S/N for 21\,cm observations with HERA (left) and SKA1-Low (right), combined with SO kSZ observations. We apply the realistic foreground filters with $m=3$. The upper left regions in each panel, above the dashed lines, are combinations of $z_0$ and $\Delta z$ that would require post-reionization contributions, which are not included in this work. We find that one can obtain the largest S/N at redshifts close to $z_0 \sim 9$ for intermediate redshift chunk sizes, $\Delta z$. Also, the forecasts are better for SKA1-Low than HERA owing to SKA1-Low's superior ability to sample small-scale modes.}
    \label{fig:SNR_single}
\end{figure*}
where we define:
\begin{equation}
    C_{\ell}^{TT,f} = f^{\rm kSZ}(\ell)^2(C_\ell^{TT}+C_\ell^{\rm kSZ, reion}+C_\ell^{\rm kSZ,late}+N_\ell)\,.
\end{equation}



Intuitively, one might estimate $C_\ell^{{\rm{21cm^221cm^2}},f}$ with the Gaussian approximation and the Limber approximation \citep{1953ApJ...117..134L,2004ApJ...606...46D} analogously. However, the 21\,cm field in the EoR is highly non-Gaussian \citep{2007MNRAS.379.1647W,2004ApJ...613...16F}, and so the Gaussian approximation is expected to be inaccurate.
Therefore, we instead choose to evaluate $C_\ell^{{\rm{21cm^221cm^2}},f}$ using the 21\,cm mock signal data with thermal noise included.\footnote{We evaluate the ratio of $C_\ell^{{\rm{21cm^221cm^2}},f}$ between the simulated data and the Gaussian approximation. In the case of HERA, the Gaussian approximation overestimates $C_\ell^{{\rm{21cm^221cm^2}},f}$ by a factor of 10-100. In the case of SKA1-Low, it overestimates $C_\ell^{{\rm{21cm^221cm^2}},f}$ by a factor of 1-10.}

In order to calculate the noise term, we need to add instrumental noise to the mock 21 cm signal data. We assume Gaussian noise and use the covariance matrix of Equations~(\ref{eqn:hera-noise}) and~(\ref{eqn:ska-noise}). 

Figure~\ref{fig:SNR_single} shows the angular cross-power spectrum of the filtered ${\rm 21cm^2}$ fields (upper panels) with a realistic filter of $m=3$ at $\ell=1600$, and the signal-to-noise-ratio (S/N) (bottom panels), as a function of $z_0$ and $\Delta z$. We adopt the fiducial reionization history and assume that the 21\,cm field is measured by HERA (left panels) and SKA1-Low (right panels). 
In both cases, the variance term $C_\ell^{{\rm{21cm^221cm^2}},f}$ decreases as the chunk size increases. This occurs because the variance decreases as additional, nearly independent, redshift slices are included in a redshift chunk. However, there is a tradeoff because the signal amplitude drops as the chunk size grows (see Figure \ref{fig:chunksize}). 

The dependencies of the variance term on the central redshift $z_0$ are different in the cases of HERA and SKA-Low. In the case of HERA, the thermal noise contribution is comparable to the sample variance contribution. At lower redshifts, the transverse co-moving distance $D_M$ is smaller. Therefore, $k_\perp$ increases at fixed $\ell$. Hence, the large thermal noise on scales with low baseline counts enters the overall convolution and this enhances the variance term $C_\ell^{{\rm{21cm^221cm^2}},f}$ dramatically. However, we do not expect the variance term to keep increasing as the $X^2(\nu)Y(\nu)$ factor in the noise spectrum should decrease with redshift and finally lower the variance term. Therefore, in the case of HERA, $C_\ell^{{\rm{21cm^221cm^2}},f}$ should first increase and then decrease as the redshift decreases. This expectation is consistent with the top left panel of Figure~\ref{fig:SNR_single}. 
In the case of SKA1-Low, the thermal noise and the sample variance are weakly dependent on frequency and multipole moment, $\ell$. Therefore, it is weakly dependent on $z_0$, as shown in the top right panel of Figure~\ref{fig:SNR_single}. 
Based on the above discussion and the fact that the ${\rm kSZ^2}$-$\rm{21cm^2}$ cross-power spectrum peaks at $z\sim 9$ in the fiducial reionization model (see Figure~\ref{fig:historydependence}), 
it is reasonable that the largest S/N is obtained at $z_0 \sim 9$ with an intermediate chunk size. Further, it makes sense that SKA1-Low achieves better sensitivity than HERA, as shown in the bottom panels of Figure~\ref{fig:SNR_single}.

\begin{figure*}[htbp]
    \centering
    \includegraphics[width=0.6\columnwidth]{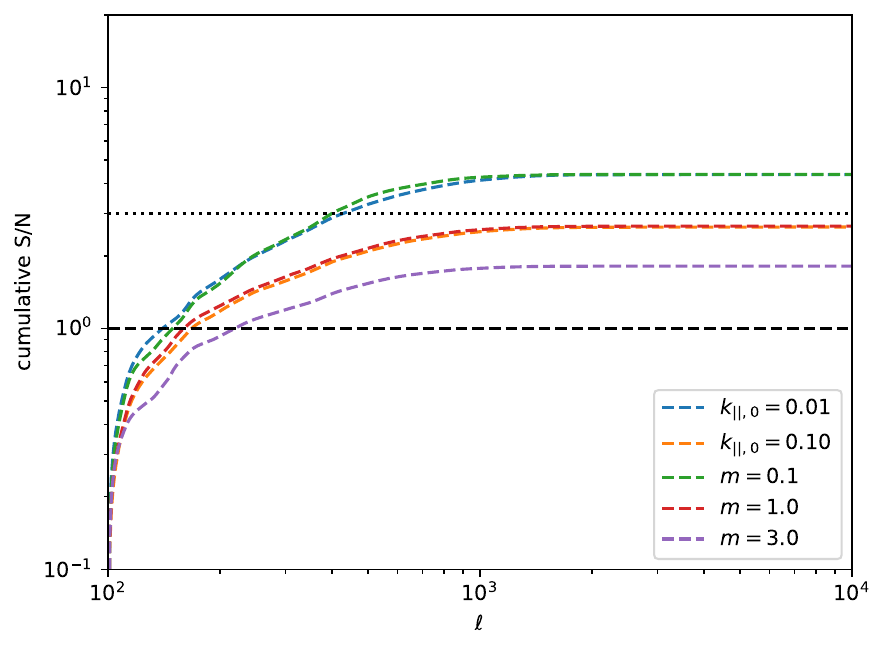}
    \includegraphics[width=0.6\columnwidth]{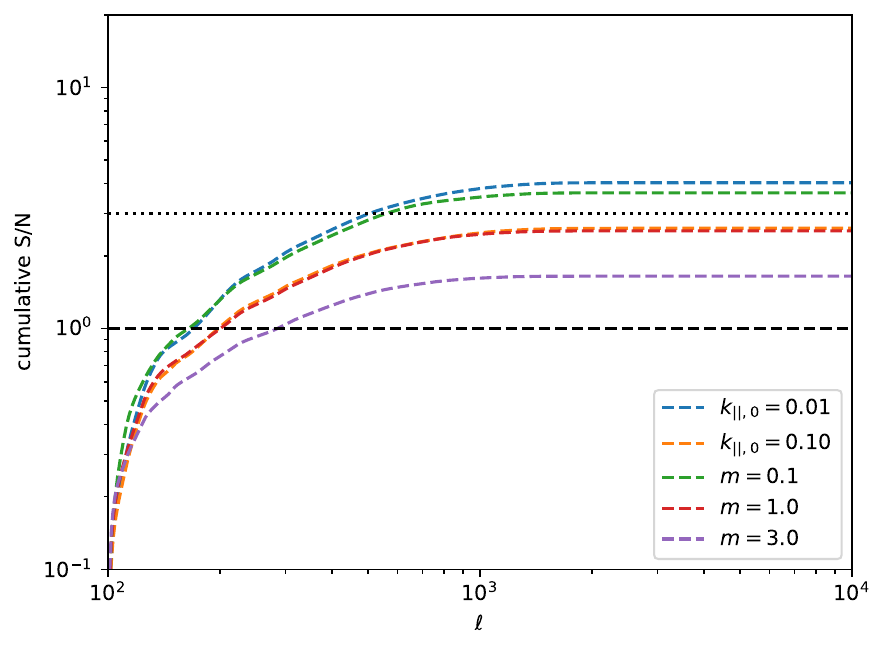}
    \includegraphics[width=0.6\columnwidth]{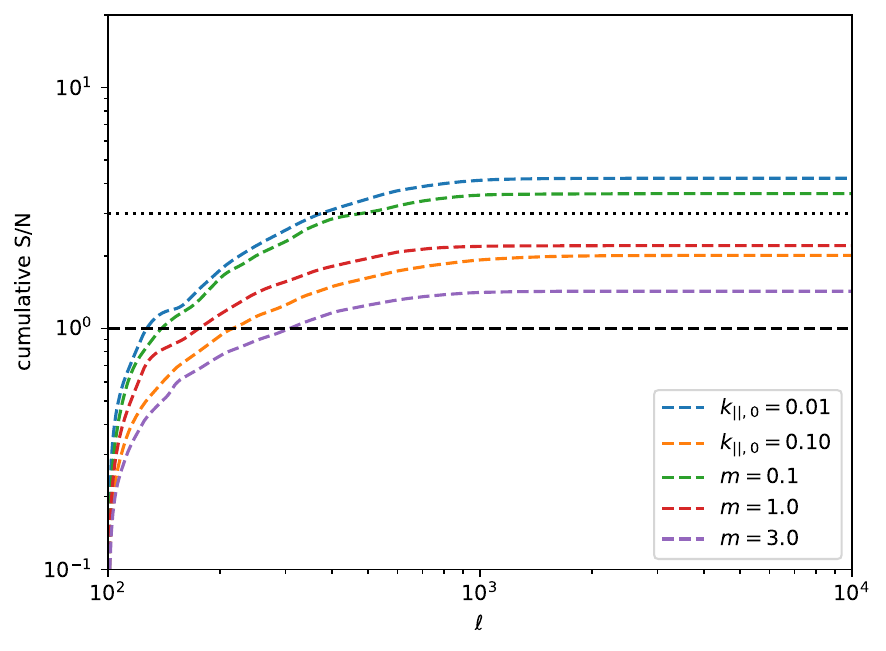}
    \caption{The cumulative S/N for various scenarios as a function of $\ell$ mode. We evaluate them from $\ell_{\rm min}=100$ to $\ell_{\rm max}=10000$ and search for the combination of $z_0$ and $\Delta z$ that maximizes the S/N. We assume an observation with HERA in combination with CMB-S4 and vary the foreground filters employed. From left to right, we consider the fiducial, late, and short reionization models, respectively.}
    \label{fig:bestsnr_hera}
\end{figure*}
\begin{figure*}[htbp]
    \centering
    \includegraphics[width=0.6\columnwidth]{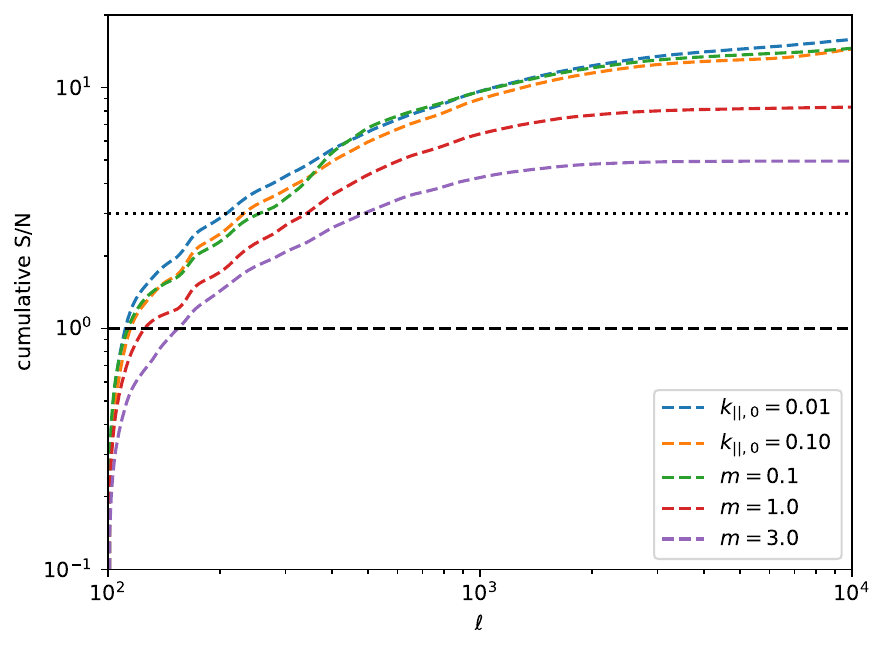}
    \includegraphics[width=0.6\columnwidth]{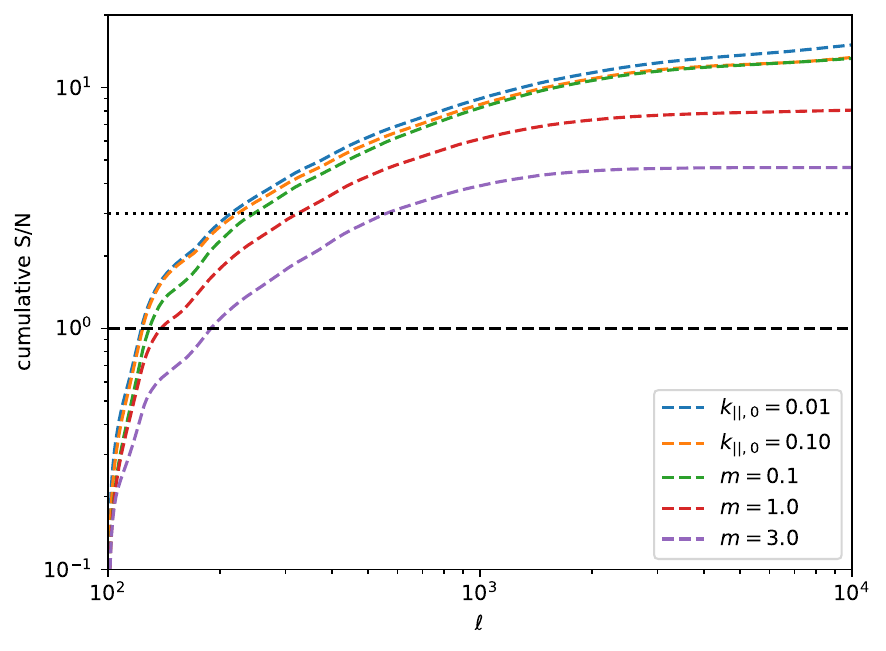}
    \includegraphics[width=0.6\columnwidth]{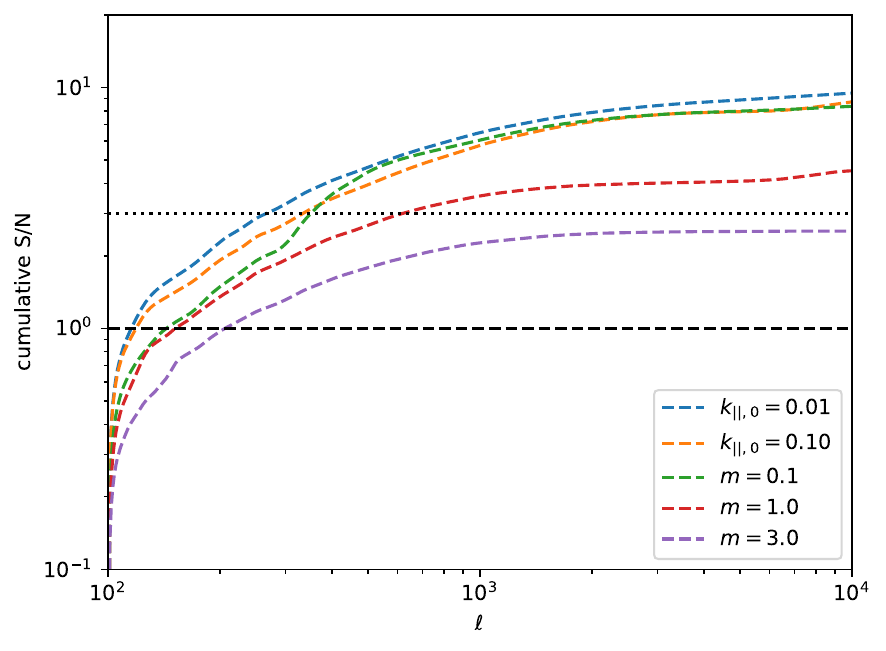}
    \caption{Same as Figure~\ref{fig:bestsnr_hera}, but assuming the 21\,cm signal as measured by SKA1-Low.}
    \label{fig:bestsnr_SKA}
\end{figure*}

We can generalize these calculations of the S/N for a single redshift bin to determine the cumulative S/N across all reionization-era redshifts. The total S/N can be written as:
\begin{equation}
    \left(\frac{\rm S}{\rm N}\right)^2_{\rm tot} = \sum_{z_0}\left(\frac{\rm S}{\rm N}\right)^2_{z_0}
\end{equation}
In Figure~\ref{fig:bestsnr_hera}, we show the optimal cumulative S/N for all scenarios with HERA in combination with CMB-S4, using different foreground filters, and reionization history models. From left to right, we consider the fiducial, late, and short reionization models respectively. In each scenario, we vary the width $\Delta z$ to find the maximum value of the S/N. We find that in the case of HERA $\times$ CMB-S4, the kSZ$^2$--21\,cm$^2$ angular cross-power spectrum is detectable at a significance larger then $1\sigma$ using a realistic $m=3$ foreground filter with $\Delta z =$ 2.25 (Fiducial), 1.62 (Late), and 1.42 (Short), respectively. If future 21\,cm signal data-processing efforts allow us to push the foreground filter to a smaller size, i.e. $k_{\parallel,0}=0.01$, the cumulative S/N will increase to $4\sigma$ with $\Delta z = $ 3.86 (Fiducial), 3.23 (Late), and 5.22 (Short), respectively. The dependence of the cumulative S/N on $\Delta z$ results from the dependencies of the S/N in a single redshift on $\Delta z$ and $z_0$. It is hard to interpret and may vary between reionization models. We also present the detectability using SKA1-Low in Figure~\ref{fig:bestsnr_SKA}. Since SKA1-Low has more small-scale baseline counts, the correlated 21\,cm signal can be measured more accurately and the optimal S/N is larger. Using a realistic $m=3$ foreground filter, we should still be able to detect the cross-power spectrum with a significance over $4\sigma$ for $\Delta z = $ 2.82 (Fiducial) and 4.23 (Late), respectively, except in the case of the short reionization model. In fact, the S/N value in the short reionization scenario is always smaller than in the other models. In the case of the fiducial and late reionization models, we find that the S/N reaches 15$\sigma$ for a filter with $k_\parallel,0 = 0.01$ and $\Delta z = 6.0$. In the case of the short reionization model, we forecast a significance of 9.48$\sigma$ with the same foreground filter and $\Delta z = 5.22$.

We also evaluate the S/N using SO and CMB-HD. We list the best S/N values of various scenarios in Table~\ref{tab:SNR-HERA} and Table~\ref{tab:SNR-SKA}, for 21\,cm observations with HERA and SKA1-Low, respectively. In the case of HERA (SKA1-Low), using a realistic $m=3$ foreground filter and a similar value of $\Delta z$, SO will downgrade the S/N to $\leq1$ (3) $\sigma$ while CMB-HD will upgrade the S/N to at least 5 (9) $\sigma$. If one uses an optimistic $k_{0,\parallel}=0.01$ foreground filter and a similar value of $\Delta z$, SO will downgrade the S/N to around 2 (over 5) $\sigma$ while CMB-HD will boost the S/N to around 15 (over 37) $\sigma$.


\section{Discussion}
The above S/N estimates help determine the optimal redshift windows for kSZ$^2$--21\,cm$^2$ cross-power spectrum measurements. Our explorations reveal several possible ways to further improve the S/N forecasts. Some axes for improving the S/N should match those discussed in \citet{2022ApJ...928..162L,2023ApJ...944...59L}, including further mitigating CMB foregrounds and increasing the sky coverage. They should help to increase the S/N here in the same way. We will not repeat similar discussions here but focus on the unique improvements with respect to our current signals.

\subsection{Long baseline counts}
Our S/N estimates assume 21 cm observations with HERA and SKA1-Low, respectively. From Figure~\ref{fig:chunksize} and Figure~\ref{fig:historydependence}, we find that SKA1-Low can better preserve the intrinsic features of the signals. SKA1-Low can also improve the S/N by a factor of 3-4 compared to HERA (see Table~\ref{tab:SNR-HERA} and Table~\ref{tab:SNR-SKA}).  
SKA1-Low performs better than HERA since it includes more long baselines at $u \gtrsim 400$ (see Figure~\ref{fig:baseline_counts}). This boosts the detectability of the cross-power spectrum signal significantly since the small-scale modes better sampled by SKA1-Low match those passing through the kSZ filter in the CMB observations. These modes contribute the most to the signal. Therefore, the S/N improves as baseline counts are added on scales where the cross-correlation signals are strong. 


\begin{table*}[htbp]
    \centering
    \caption{The best cumulative S/N value, assuming 21\,cm to be measured by HERA.}
    \begin{threeparttable}
    \begin{tabular}{ccccccc}
    \hline
       \hline
       CMB Experiment& History& {$k_{\parallel,0}=0.01$}&{$k_{\parallel,0}=0.10$}&{$m=0.1$}&{$m=1.0$}&{$m=3.0$}\\
       \hline
      CMB-SO &Fiducial&2.34&1.40& 2.33& 1.42&0.98\\
       &Late&2.18& 1.39&1.96 &1.36 &0.88\\
       &Short&2.35 &1.11&2.02&1.23 &0.80\\
       \hline
      CMB-S4 &Fiducial&4.35 &2.64&4.36&2.66 &1.81\\
       &Late&4.02 &2.60&3.65&2.54&1.64\\
       &Short&4.20&2.01 &3.63 &2.21 &1.43\\
       \hline
       CMB-HD &Fiducial&16.03&9.92&16.22 &9.96&19.32\tnote{*}\\
       &Late&15.07&9.80&13.74&9.53&6.23\\
       &Short&14.83&7.28 &12.91&7.80&5.11\\
       \hline
    \end{tabular}
    \begin{tablenotes}
    \footnotesize
    \item[*] In this case, the S/N is mainly limited by the 21\,cm instrumental noise. A large foreground filter size can remove more noise power in the high $k_\perp$ regions and produce a large S/N value.
    \end{tablenotes}
    \end{threeparttable}
    \label{tab:SNR-HERA}
\end{table*}


\begin{table*}[htbp]
    \centering
    \caption{The best cumulative S/N value, assuming 21\,cm to be measured by SKA1-Low.}
    \begin{tabular}{ccccccc}
    \hline
       \hline
       CMB Experiment& History&{$k_{\parallel,0}=0.01$}&{$k_{\parallel,0}=0.10$}&{$m=0.1$}&{$m=1.0$}&{$m=3.0$}\\
       \hline
      CMB-SO &Fiducial&7.96&7.20&7.32&4.24 &2.56\\
       &Late&7.50&6.60&6.62&4.12&2.38\\
       &Short&5.04&4.56&4.48&2.53&1.38\\
       \hline
      CMB-S4 &Fiducial&15.85&14.46&14.55 &8.31&4.95\\
       &Late&15.02&13.33&13.21&8.05&4.65\\
       &Short&9.48&8.72&8.36&4.51&2.54\\
       \hline
       CMB-HD &Fiducial&66.78&60.64&60.37 &34.13&19.95\\
       &Late&65.05&58.67&56.76&33.36&19.32\\
       &Short&37.17&34.38&32.05&15.85 &9.60\\
       \hline
    \end{tabular}
    \label{tab:SNR-SKA}
\end{table*}

\subsection{Recovery of lost modes}
The S/N forecasts may also be improved if some of the long-wavelength radial modes in the 21 cm data may be recovered. Figure~\ref{fig:filtersize} has shown the significant impacts of the filter size on the signals. A large foreground filter size could make the signal less smooth and reduce the amplitude of the signal. We also find that in most cases, the best cumulative S/N value decreases as the filter size increases. If one can partially remove the foregrounds and use a highpass filter with $k_{\parallel,0}=0.01$ instead of a wedge filter with $m=3$, the cumulative S/N will increase by a factor of 2-3. The recovery of such modes should be feasible because different Fourier modes are coupled due to cosmic structure formation (as our statistic here also exploits). Indeed, recent work developed successful methods based on tidal reconstruction \citep{2018PhRvD..98d3511Z} and machine learning \citep{2024MNRAS.529.3684K,2024arXiv240721097S} to recover modes lost in the foreground removal process.

\section{Conclusion}\label{sec: conclusion}
We have simulated the $\rm{kSZ^2}$-$\rm{21cm^2}$ cross-power spectrum during the EoR. We apply filters to each signal, before squaring, to mimic mitigation steps that will be necessary with real data to reduce contamination from CMB foregrounds, including the late-time kSZ signal, along with 21\,cm foregrounds and instrumental noise in both the CMB and 21\,cm data. We then square the kSZ and 21\,cm signals from the EoR to avoid large-scale cancellations and compensate for the loss of low $k_{||}$ 21\,cm modes, respectively. We find that the simulated cross-power spectrum shows generic evolutionary trends as reionization proceeds, at least across the three disparate reionization histories explored here. It peaks for all modes around $400<\ell<1600$ when the global ionization fraction is around $x_{\rm HII} \sim 0.15$, while the signal reaches a minimum value around the mid-point of reionization (near $x_{\rm HII} \sim 0.5$).  
The 21\,cm filter weakly influences the generic features of our signal.

Note that the behavior found in this work may be somewhat model dependent, as all of our models employ the {\tt zreion} framework.
In future forecasts, it may be interesting to explore uncertainties in the astrophysical modeling and semi-numeric/numeric methods \citep{2022PASP..134d4001Z}. For example, one might compare with {\tt 21cmFAST\footnote{\url{https://github.com/21cmfast/21cmFAST}}} simulations \citep{2011MNRAS.411..955M,2020JOSS....5.2582M}.  

We estimate the detectability in different combinations of surveys and for different foreground filters. We forecast that cross-correlation measurements from CMB-S4 and SKA1-Low data can achieve $4\sigma$ detections. Using CMB-HD can increase the significance to $\sim 9-20\sigma$, depending on the reionization model. We find that cross-correlations with HERA are less sensitive, due to the lack of long baselines in HERA, but we nevertheless expect a $1\sigma$ detection for CMB-S4 and a $\sim 5-20\sigma$ for CMB-HD, depending on the reionization model.

In principle, the analysis method applied in this work could be extended to other large-scale tracers during the EoR, e.g. using galaxy samples and other emission lines with related foreground contamination concerns. In this work, we focus on the case of the $\rm{kSZ^2}$-$\rm{21cm^2}$ cross-power spectrum because our current understanding of 21\,cm foregrounds are more advanced, and given that SKA1-Low is expected to be taking data along side next-generation CMB measurements, such as CMB-S4. Our forecasts suggest that this combination will allow detections of the $\rm{kSZ^2}$-$\rm{21cm^2}$ cross-power spectrum statistic explored here. 

\section*{Acknowledgements}
We thank James Aguirre, Abinash Kumar Shaw, Clinton Stevens, Jason Sun, and Jianrong Tan for useful discussions. PL is supported by NSF award number 2206602, NASA award number 80NSSC22K0818, and the Simons Foundation award number 00007127. YM is supported by the National SKA Program of China (grant No. 2020SKA0110401). YZM is supported by the National Research Foundation of South Africa (grants No. 150580, and No. CHN22111069370). This work used the Extreme Science and Engineering Discovery Environment (XSEDE), which is supported by National Science Foundation grant number ACI-1548562 \citep{2014CSE....16e..62T}. Specifically, it used the Bridges-2 system, which is supported by NSF award number ACI-1445606, at the Pittsburgh Supercomputing Center \citep[PSC, ][]{10.1145/2792745.2792775}. We also acknowledge the Tsinghua Astrophysics High-Performance Computing platform at Tsinghua University for providing computational and data storage resources that have contributed to part of the research results reported within this paper.

\begin{appendix}
\section{Analytical interpretation of squaring the 21\,cm fields}
\label{appendix_a}
In this section, we use the convolution theorem to explain why the squared 21\,cm map is non-vanishing even after all of the low $k_\parallel$ modes are removed. For convenience, we define $\tilde{A}(\mathbf{k})$ as the Fourier transform of the filtered and then squared 21\,cm field. 
According to the convolution theorem, the Fourier transform of a product of two functions is the convolution of their Fourier transform. Therefore, for $k_\parallel=0$, $\tilde{A}(k_\parallel=0,\mathbf{k}_\perp)$ can be written as:
\begin{eqnarray}
    \tilde{A}(k_\parallel=0,\mathbf{k}_\perp) &=& \int \frac{d^3\mathbf{k}^\prime}{(2\pi)^3}\nonumber\\
    \Delta T^{\rm 21cm}_f(-k_\parallel^\prime,\mathbf{k}_\perp&-&\mathbf{k}_\perp^\prime)\Delta T^{\rm 21cm}_f(k_\parallel^\prime,\mathbf{k}_\perp^\prime)
\end{eqnarray}

Note that $\tilde{A}(k_\parallel=0,\mathbf{k}_\perp)$ is non-vanishing since high $k_\parallel^\prime$ $\Delta T^{\rm 21cm}_f$ modes should inevitably contribute to this convolution.

\section{$C_\ell^{\rm kSZ^2\times\rm 21cm^2}$ in Limber Approximation}
\label{appendix_b}
In this section, we relate $C_\ell^{\rm kSZ^2\times\rm 21cm^2}$ to an underlying trispectrum statistic to help reveal the physical content of this statistic. Following the derivations in \citet{2016PhRvD..94l3526F,2022ApJ...928..162L}, we can apply the Limber approximation and write the $C_\ell^{\rm kSZ^2\times\rm 21cm^2}$ statistic as:
\begin{widetext}
\begin{eqnarray}
    C_\ell^{\rm kSZ^2\times\rm 21cm^2} &=& \int \frac{d\chi}{\chi^6}W(\chi)g^2(\chi)\int \frac{d^2\boldsymbol{L}_1}{(2\pi)^2}\frac{dq_\parallel}{2\pi}\frac{d^2\boldsymbol{L}_2}{(2\pi)^2}f^{\rm 21cm}\left(-q_\parallel,\frac{\boldsymbol{\ell}-\boldsymbol{L}_1}{\chi}\right)f^{\rm 21cm}\left(q_\parallel,\frac{\boldsymbol{L}_1}{\chi}\right)\nonumber\\
    f^{\rm kSZ}(-\boldsymbol{\ell}&-&\boldsymbol{L}_2)f^{\rm kSZ}(\boldsymbol{L}_2)T^{21{\rm cm},\ 21{\rm cm},\ p_{\hat{n}},\ p_{\hat{n}}}\Bigg[\left(-q_\parallel,\frac{\boldsymbol{\ell}-\boldsymbol{L}_1}{\chi}\right);\left(q_\parallel,\frac{\boldsymbol{L}_1}{\chi}\right);\left(0,-\boldsymbol{\ell}-\boldsymbol{L}_2\right);\left(0,\boldsymbol{L}_2\right)\Bigg]\,,
\end{eqnarray}
\end{widetext}
where the trispectrum above, $T^{21{\rm cm},\ 21{\rm cm},\ p_{\hat{n}},\ p_{\hat{n}}}$, involves the unfiltered 21\,cm fluctuations at two different wavenumbers and the spatial variations in the line-of-sight component of the electron momentum fields, $p_{\hat{n}}$, at two different multipoles. Although this expression is not explicitly evaluated in this work, it nevertheless demonstrates that $C_\ell^{\rm kSZ^2\times\rm 21cm^2}$ probes an integral over the trispectrum.
The contribution of different kSZ and 21\,cm modes to this integral are determined, in part, by the filters applied. 

\end{appendix}
%





\bibliography{kszsq-21cmsq-xcorr}{}
\bibliographystyle{aasjournal}



\end{document}